\documentclass{jfm}
\usepackage{subfiles}
\usepackage{graphicx}
\usepackage{epstopdf,epsfig,subfloat,subfig,floatrow}
\usepackage{newtxtext}
\usepackage{newtxmath}
\usepackage{natbib}
\usepackage{amsfonts}
\usepackage{amsmath}
\usepackage{amssymb,bm,mathtools}
\usepackage{color}
\usepackage{hyperref}
\usepackage{multirow,makecell,array,booktabs}
\usepackage{soul,cancel}
\usepackage{ulem}
\usepackage{lineno}

\hypersetup{
    colorlinks = true,
    urlcolor   = blue,
    citecolor  = blue,
}

\newcommand{\RomanNumeralCaps}[1]

\allowdisplaybreaks[4]

\definecolor{orange}{RGB}{255,100,0}

\title{Lateral turbulent jet in rarefied environment}
\author{Songyan Tian,
      Lei Wu%\aff{1}
        \corresp{\email{wul@sustech.edu.cn}}
        \and  Minping Wan
        }

\affiliation{
  Department of Mechanics and Aerospace Engineering, Southern University of Science and Technology, Shenzhen 518055, China
  }

\begin{document}
\maketitle

\begin{abstract}
Lateral jets play a crucial role in controlling the trajectory and aerodynamic heating of hypersonic vehicles. However, the complex interaction between turbulent and rarefaction effects has rarely been examined. This study fills this knowledge gap by employing the newly developed GSIS-SST method [J. Fluid Mech. 1002 (2025) A10], which combines the shear stress transport (SST) model for turbulent flow and the general synthetic iterative scheme (GSIS) for rarefied gas flow. It is found that, at altitudes from 50~km to 80~km, the maximum relative difference in the pitch moment between the GSIS-SST and pure GSIS (SST) reaches 28\% (20\%). While the jet is supposed to reduce the surface heat flux, its turbulence significantly diminishes this reduction, e.g., the GSIS-SST predicts a heat flux about one order of magnitude higher than the GSIS when the jet pressure ratio is 1.5. Increasing the angle of attack intensifies local turbulence, resulting in expanded discrepancies in shear stress and heat flux between GSIS-SST and GSIS. These insights enhance our comprehension of lateral jet flows and highlight the importance of accounting for both turbulent and rarefaction effects in medium-altitude hypersonic flight. 
\end{abstract}

% !TeX spellcheck = en_US

\section{Introduction}

The persistent drive for increased velocity and enhanced maneuverability of hypersonic vehicles is transforming the landscape of aerospace technology. To accomplish high-speed, high-altitude maneuvers, these vehicles must contend with diminished dynamic pressure and extreme aerodynamic heating that outstrip the capabilities of conventional flight control systems. Active flow controls, such as lateral jets, have become a critical technology to mitigate the detrimental impacts of these extreme conditions and sustain control in hypersonic flights~\citep{Mahesh2013Interaction}.

At low altitudes, the Navier-Stokes (NS) equations describe the continuum gas flows. Turbulence is expected due to the high Reynolds numbers (Re), and the numerical methods such as Reynolds-Averaged Navier-Stokes (RANS) method and large eddy simulations~\citep{Boles2010Large,Miller2018Transient,Sanaka2024Re} are used to study the dynamic interaction between the jet and the free stream.

At high altitudes, rarefaction effects gain prominence as the Knudsen number (Kn) increases\footnote{The Knudsen number is defined as the mean free path of gas molecular over the characteristic length $L$. If the same length $L$ is used in the definition of the Reynolds number, then $\text{Kn}\propto\text{Ma}/\text{Re}$.}, which not only lead to the velocity slip and temperature jump at solid walls but also alter the constitutive relations. Consequently, the continuum hypothesis underlying the NS equations is no longer valid, necessitating the use of the Boltzmann equation. The Boltzmann equation, which is defined in the six-dimensional phase space, is usually solved by the direct simulation Monte Carlo (DSMC) method \citep{Bird1994Molecular, Boyd1995Predicting} and the discrete velocity method~\citep{Aristov2001Direct,Xu2010unified,Liu2024Further}.

Theoretically, the Boltzmann equation provides an accurate model for dilute gas flows from the continuum to free-molecular regimes. The NS equations, which are effective in the continuum flow regime where the Knudsen number is small, can be derived from the Boltzmann equation through the Chapman-Enskog expansion~\citep{Chapman1990Mathematical}. 
However, directly solving the Boltzmann equation for continuum flows, particularly in the presence of turbulence, is computationally intensive. For instance, simulating the turbulent Taylor-Green vortex using DSMC required over half a million CPU cores for 500 hours~\citep{Gallis2017Molecular}. 
The DSMC method is computationally intensive because the splitting of the streaming and collision operators in the Boltzmann equation requires the spatial cell size and time step to be smaller than the molecular mean free path and mean collision time, respectively. 
Although various kinetic methods have been developed~\citep{Xu2010unified,Gorji2011JFM,Liu2020Unified,Su2020Can,Fei2023JCP} to provide faster computation times than DSMC in near-continuum flow regimes, they still face challenges in efficiently handling turbulence.
Therefore, for lateral jets in hypersonic flows, most studies focus on either high-density jet flows modeled with continuum solvers~\citep{Rowton2024ANumerical} or low-density rarefied flows modeled with kinetic solvers~\citep{Gimelshein2002Modeling,Karpuzcu2023Study}. % ,Zhao2023Numerical

However, at intermediate altitudes, a new realm of fluid mechanics emerges: the coexistence and interaction of turbulence and rarefied gases. If the turbulent and rarefied flows occupy a small and large spatial domain respectively, they possess comparable energies, and their interplay may result in substantial changes in macroscopic flow fields. In such circumstance, 
continuum solvers fail to account for rarefaction effects, and kinetic solvers face prohibitive computational costs when turbulence is involved. 
To overcome these challenges, we have recently introduced the GSIS-SST solver~\citep{Tian2024Multiscale}, which integrates the general synthetic iterative scheme (GSIS) for the Boltzmann equation of rarefied gas flows~\citep{Su2020Can,Su2020Fast} with the $k$-$\omega$ shear stress transport (SST) model for turbulent flows \citep{Menter1994Two}. The GSIS provides fast convergence and asymptotic-preserving properties, allowing for efficient solutions of the Boltzmann equation on coarse grids, while the SST model captures turbulent effects and alleviates the stringent grid resolution requirements. Together, the GSIS-SST offers a cohesive description spanning from rarefied gas flow to turbulent flow. 

With the GSIS-SST solver, \cite{Tian2024Multiscale} have revealed that the interaction between an opposing turbulent jet and hypersonic rarefied gas flow substantially alters the  heat flux on hypersonic vehicles. In this paper, we shall explore the interaction between lateral turbulent jets and rarefied hypersonic flows.
%The remainder of this paper is organized as follows: Section \ref{sec:geom&case} elaborates on the geometric model and evaluates the asymptotic properties of GSIS-SST; Sections \ref{sec:PR} and \ref{sec:AoA} highlight the role of turbulence in rarefied environments by comparing critical flow fields between GSIS-SST and GSIS; finally, conclusions are presented in Section \ref{sec:conclusion}.

%With the GSIS-SST, we aim to address the knowledge gap by investigating the influence of lateral turbulent jets at high altitudes and providing new insights into the influence of rarefaction, non-equilibrium and jet-induced turbulence on hypersonic flow dynamics.
% We focus on key flow structures and surface quantities, such as the interaction between the jet and free stream, the aerodynamic force and surface heat flux.

%\input{Methodology}

\section{The GSIS-SST and its asymptotic behavior}\label{sec:geom&case}

\begin{figure}
	\centering
	{\includegraphics[width=0.355\linewidth,trim={30 -30 210 220}, clip]{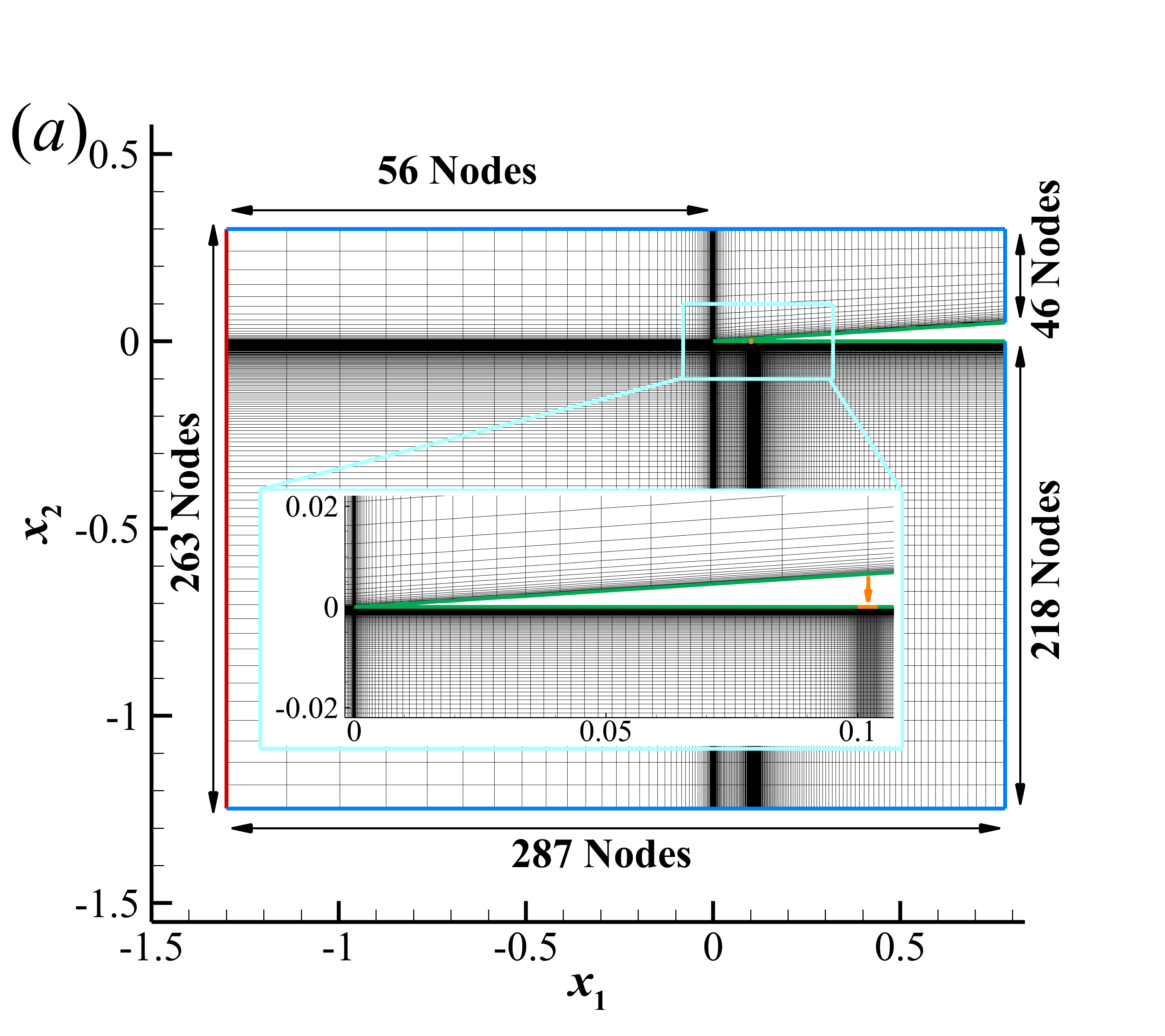}} 
        {\includegraphics[width=0.635\linewidth, trim={50 40 230 0}, clip]{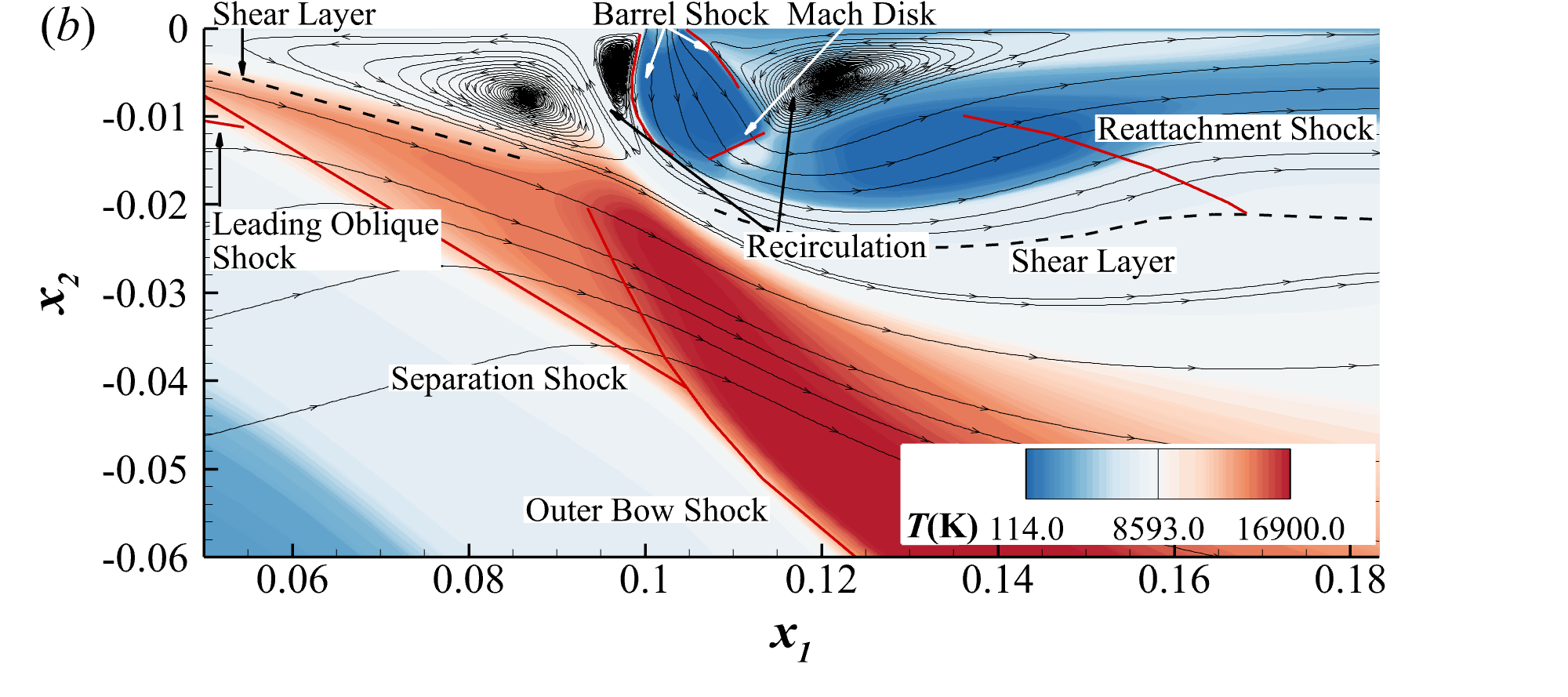}}
	\caption{(a) Geometry of the sharp leading edge and spatial grid. The free flow boundary is indicated in red, the outflow boundary is marked in blue, the wall is colored green, and the jet, situated approximately at $x_1=0.1~\text{m}$, is highlighted in orange. 
    In the spatial discretization, there is 67,732 cells in total. Specifically, 84 nodes are positioned before the jet and 120 nodes after it on the lower horizontal surface, while the jet exit is allocated 30 nodes. The height of the first grid layer is one micron.
    % with 263 nodes for free flow boundary, 56 nodes assigned in front of the nose-tip, 84 nodes positioned before the jet and 120 nodes after it on the lower horizontal surface, while the jet exit is allocated 30 nodes. The height of the first grid layer is one micron. 
    (b) Contour of the total temperature and streamlines of the lateral jet at altitude 70~km with $\text{PR}=2.5$ and angle of attack (AoA)=15° from GSIS-SST. Shear layers are represented by dashed lines, while strong shocks are outlined by solid red lines. According to the U.S. Standard Atmosphere~\citep{United1976United}, the free flow pressure is $p=5.22$ Pa, temperature is $T=219.6$~K, viscosity is $\mu=1.44\times 10^{-5}~\text{Pa}\cdot \text{s}$, so the Knudsen number is Kn=$\frac{\mu}{pL}\sqrt{{\pi RT}/{2}}=0.00866$ and Reynolds numbers is 4280. }
\label{fig:H70TtComment}
\end{figure}

As shown in figure~\ref{fig:H70TtComment}, a hypersonic flow of nitrogen gas with an incoming Mach number of 25 is considered over a sharp leading edge (SLE)\footnote{We focus on the lower surface where the jet locates, SLE is referred to the lower flat surface hereafter.}. The total length of the leading edge is $\ell=0.78$ meters, and its thickness at the base is 0.05 meters. 
By choosing the characteristic flow length $L$ to be 0.1 m, the free stream Knudsen numbers from the altitude of 30 km to 100 km range from $3.9\times10^{-5}$ to 1.20, while the Reynolds number ranges from 94,100 to 31.

% \begin{table}
%     \centering
%     \begin{tabular}{cccccc}
%     \hline
%          Altitude (km) &$p$ (Pa) &$T$ (K) &$\mu$ ($\text{Pa}\cdot \text{s}$) & Kn=$\frac{\mu}{pL}\sqrt{\frac{\pi RT}{2}}$& Re=$\frac{p U L}{RT\mu}$\\
%          \hline
%          100    &  0.03 &   194.0  &1.29E-05
% &  1.20E+00 & 3.09E+01\\
%          90     & 0.18  &  188.0   &1.26E-05
% & 2.00E-01  &1.86E+02\\
%          80     & 1.05  &  198.6   &1.32E-05
% & 3.75E-02  &9.87E+02\\
%          75     & 2.39  &  208.4   &1.38E-05
% & 1.77E-02  &2.10E+03\\
%          70     & 5.22  &  219.6   &1.44E-05
% & 8.66E-03  &4.28E+03\\
%          65     & 10.93 &  233.3   &1.51E-05
% & 4.49E-03  &8.27E+03\\
%          60     & 21.96 &  247.0   &1.58E-05
% & 2.41E-03  &1.54E+04\\
%          50     & 79.78 &  270.7   &1.70E-05
% & 7.46E-04  &4.97E+04\\
%          40     & 287.14    &  250.4   &1.60E-05
% & 1.87E-04  &1.98E+05\\
%          30     & 1197.03   &  226.5   &1.48E-05
% & 3.94E-05  &9.41E+05\\
% \hline
%     \end{tabular}
%     \caption{
%     Free flow boundary conditions. For altitude below 90~km, data are calculated based on the U.S. standard atmosphere~\citep{United1976United}; otherwise data are extracted from \cite{Moss2006Orion}. $\mu$ is the shear viscosity, $R$ is the gas constant, $T$ is the free flow temperature, and $U$ is the incoming velocity. The characteristic flow length is $L=0.1$~m. 
%     }
%     \label{tab:FreeBC}
% \end{table}

% \leir{no heat flux control??}
This setup simulates the jet-based active flow control system for hypersonic vehicles. The exit temperature of the jet is 250 K. The central axis of the jet nozzle is positioned 0.102 m from the nose-tip. The diameter of the jet nozzle is 0.004 m. The jet is sourced from a pressurized nitrogen tank, which is characterized by the jet pressure ratio (PR). This ratio represents the ratio of the total pressure between the jet and the free flow (after the normal shock). At an altitude of 70 km, the jet exit pressure is approximately 1,000 times greater than the free stream pressure when PR=2.5; this means that the mean free path of the jet (and hence the Knudsen number) is about 1000 times smaller than that in the incoming flow. Hence the jet is assumed to have a turbulence intensity of 3\%, and the jet diameter is assigned as the turbulent length scale for this internal flow.
For the external flow, the free-stream turbulence is initialized with a turbulence intensity of 0.3\% and a fixed turbulent-to-laminar viscosity ratio $\mu_{r}=\mu_{turb}/\mu_{lam}=15$, where the subscripts $turb$ and $lam$ represent turbulent and laminar viscosities, respectively. Details about the initialization are given by \cite{Goldberg1997wall}.

%Subsequently, the turbulence kinetic energy $k$ and specific dissipation rate $\omega$ are calculated as $k=1.5(I_t U_{\infty})^2$ and $\omega=\rho_{\infty} k /\mu_{turb}$. 

% and \leir{a characteristic length-scale $l_t =\sqrt{\pi r^2}$ ($r$ is the jet nozzle radius, non-dimensionalized with reference length) is assigned to the jet. The specific dissipation rate $\omega$ for jet are calculated as $\omega=\sqrt{k}/l_t$.}

In the GSIS-SST method~\citep{Tian2024Multiscale}, the Boltzmann kinetic equation~\citep{Wu2015kinetic,li2021uncertainty} is solved together with the SST turbulence model. 
The two-dimensional  molecular velocity space $[-90,90]\times[-70,70]$, which is normalized by  $\sqrt{RT}$,  is discretized by $360 \times 280$ uniform cells~\citep{Zhang2024Efficient}.
The spatial grid independence study is conducted at an altitude of 70 km with AoA=15°, and eventually 67,732 cells are used.

Figure~\ref{fig:H70TtComment} shows the typical flow structures. The underexpanded jet emerges from the nozzle, detaches from the wall, and expands vigorously, creating an expansion fan. The left flank of the expansion fan is skewed by the incoming free flow from the left. The left barrel shock also merges with the two strong shear layers: one between the large upstream vortex and the free flow, and the other between the jet flow and the free flow. 
At approximately $x_2=-0.012~\text{m}$, the Mach disk of the jet is observed. Around $x_1=0.095~\text{m}$ and $x_2=-0.02~\text{m}$, a downward-extending bow shock is visible. Adjacent to the outer bow shock, the leading oblique shock originating from the tip of the sharp leading edge is also discernible. The free flow impinges on the jet, creating adverse pressure gradient in the boundary layer, leading to upstream flow separation, forming separation shock and shear layer between the oblique shock and the wall. Downstream the jet, a large vortex is visible, signifying a low-pressure area that draws the separated jet flow back toward the wall. As the separated jet flow reattaches to the wall, it suffers from compression, resulting in a peak of heat flux on the surface, see figure~\ref{fig:H70tauQFM} below.

% \begin{figure}[h]
% 	\centering
%    {\includegraphics[width=0.9\linewidth, trim={80 70 0 0}, clip]{Artworks/H70TtComment2.png}}
% 	\caption{Contour of the translational temperature and streamlines of the lateral jet at altitude 70~km with $\text{PR}=2.5$ and angle of attack (AoA)=15° from GSIS-SST. Shear layers are represented by dashed white lines, while strong shocks are outlined by solid red lines.}
% \label{fig:jetflowfield}
% \end{figure}

%\subsection{Asymptotic behaviors}

%The rarefied hypersonic flow field with a turbulent lateral jet features an intricate interplay of shocks, shear layers, expansion zones, and separation regions. These phenomena, coupled with localized rarefaction effects and turbulence, significantly influence the flow field and aerodynamic performance. Conventional continuum or rarefied solvers are unable to simultaneously account for all these complexities. The GSIS-SST method, specifically designed for such multiscale problems, addresses this gap effectively.

\begin{figure}
    \centering
    {\includegraphics[width=0.52\linewidth,trim={10 10 10 10}, clip]{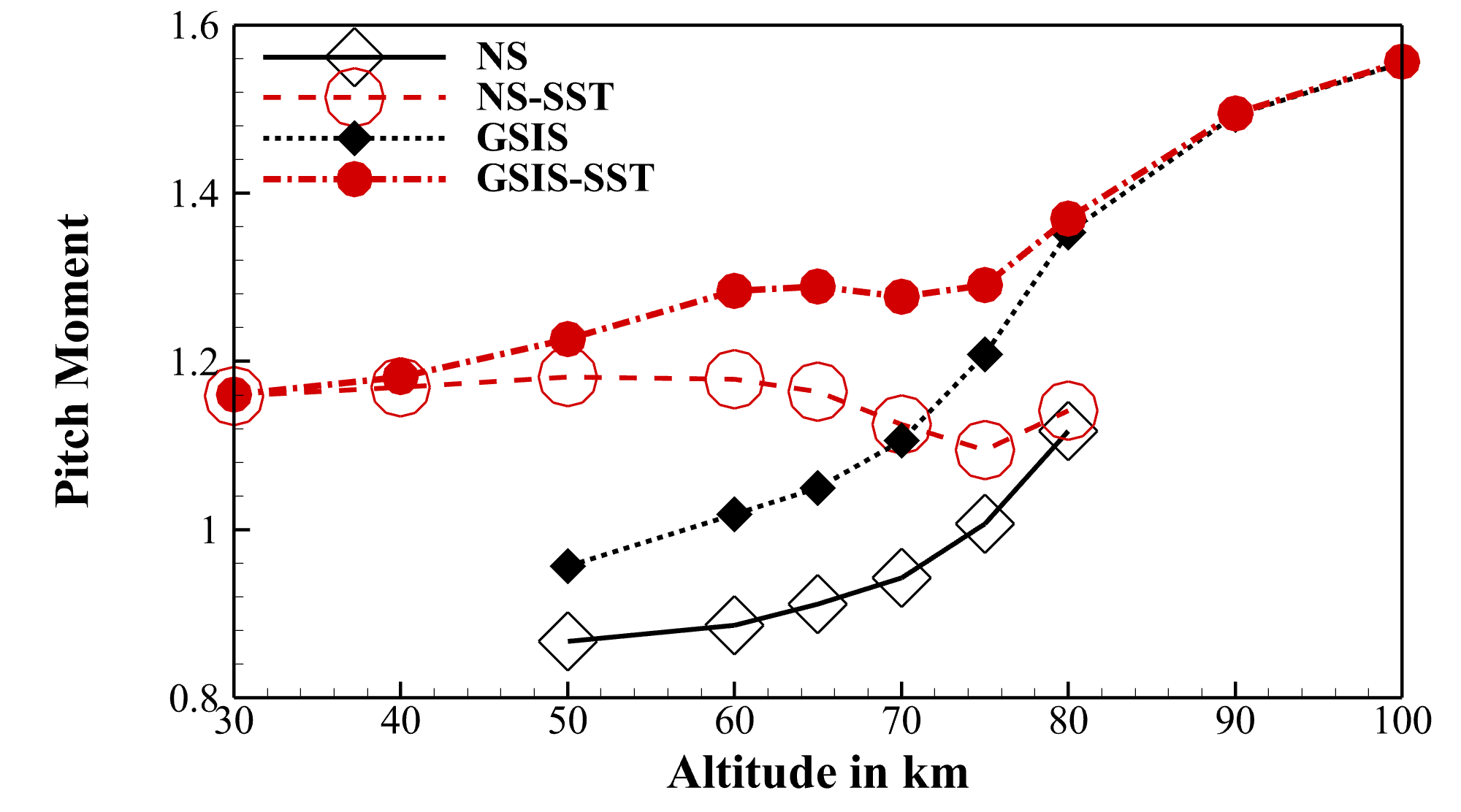}}
    \caption{Integrated pitch moment (normalized by $0.5\gamma p\text{Ma}^2\ell{}L$, $\gamma$ is the specific heat capacity ratio) on the lower surface of the model when AoA=15° and $\text{PR}=2.5$.% $0.5\gamma p{Ma}^2$ being the corresponding free flow dynamic pressure 
    %In the laminar fails region (yellow background), the solvers without turbulence modeling capability fail to provide converged steady state solutions, while in the NS fails region (blue background) the Navier-Stokes equations based continuum solvers fail to provide reasonable solutions. %(b) The relative difference of pitch moment between different pairs of solvers. The definition of relative difference for solver pair A/B is $(A/B-1)$.
    }
    \label{fig:Mx3}
\end{figure}

Figure~\ref{fig:Mx3} compares the pitch moment (integrating the product of normal stress and distance to the nose-tip at the $x_1=0~\text{m}$ over the lower surface)  between the GSIS-SST and the two-temperature NS solver, the NS solver with the same turbulence model (NS-SST), and the GSIS solver for the Boltzmann equation~\citep{Tian2024Multiscale}. 
Three asymptotic behaviors are observed. Firstly, the GSIS-SST curve coincides with the NS-SST curve at 30~km, but diverges as altitude increases, underscoring the increasing impact of rarefaction effects. 
Secondly, the GSIS-SST and GSIS curves converge at 80~km, indicating the diminishing influence of jet turbulence in low-density conditions and showcasing the adaptive capability of GSIS-SST in reducing turbulence modeling in such environments. Thirdly, for the same reason, the NS-SST and NS curves converge at 80~km. 
Finally, the GSIS curve tends toward the NS curve at lower altitudes; however, both solvers (intrinsically they are solvers of the direct numerical simulations) struggle to reach the steady state due to the turbulence, since the grid resolution is insufficient to capture the turbulence effects.  
In contrast, the GSIS-SST and NS-SST mitigate the constraints on grid resolution by integrating turbulence models \citep{Menter1994Two}, which enables the use of coarser grids while still effectively capturing the critical dynamics of the flow.

The GSIS-SST solver, which captures the two asymptotic behaviors at low and high altitudes~\citep{Su2020Fast, Tian2024Multiscale}, predicts new flow mechanics at intermediate altitudes. That is, between 50~km and 80~km, the highest relative difference in the pitch moment between the GSIS-SST and GSIS/NS/NS-SST reaches about 28\%/45\%/20\%, respectively. This underscores the synergistic effect of turbulence and rarefaction, which will be explored below.

\section{The influence of jet PR}\label{sec:PR}

Three jet PRs (i.e., PR=1.5, 2.5, and 3.5) are considered at AoA=15° and an altitude of 70 km, at which both rarefaction and turbulence effects are significant. As the PR is directly related to the density of the jet flow, and consequently the turbulent viscosity, varying $\text{PR}$ enables a comprehensive analysis of turbulence's impact on the flow field across varying intensities. 

\subsection{Turbulence influence on surface quantities}

\begin{figure}
    \centering
    {\includegraphics[width=0.32\linewidth]{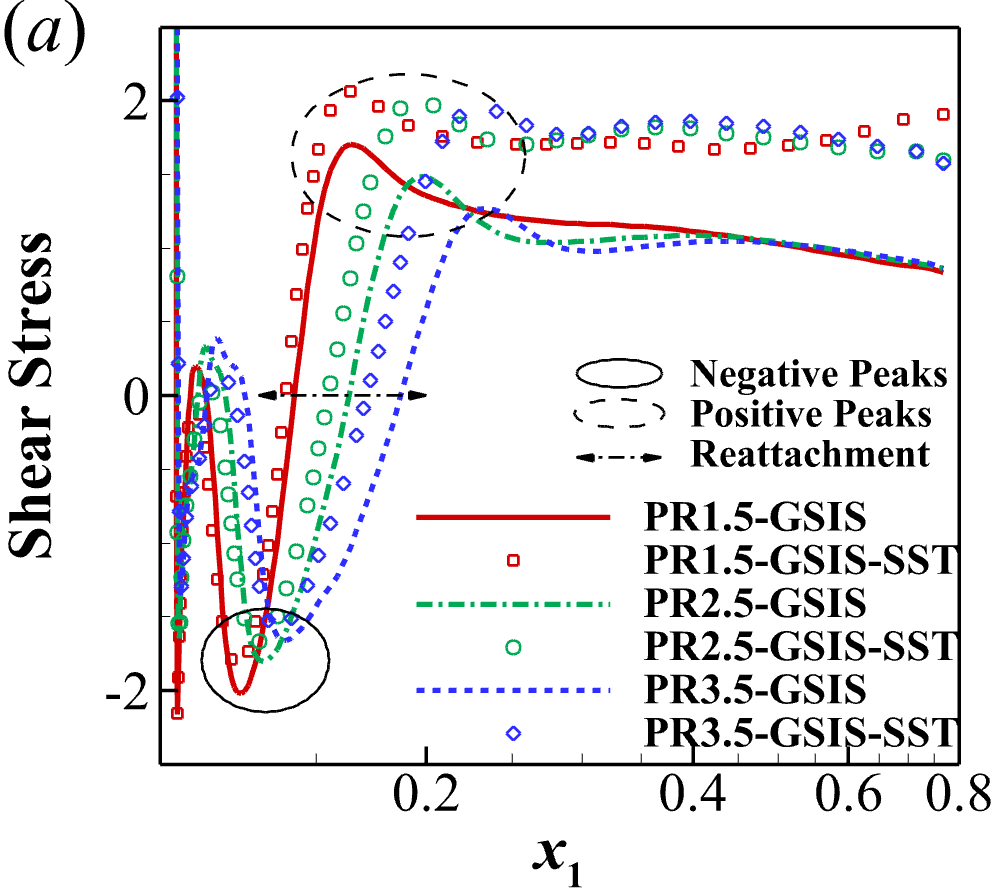}}
    {\includegraphics[width=0.32\linewidth]{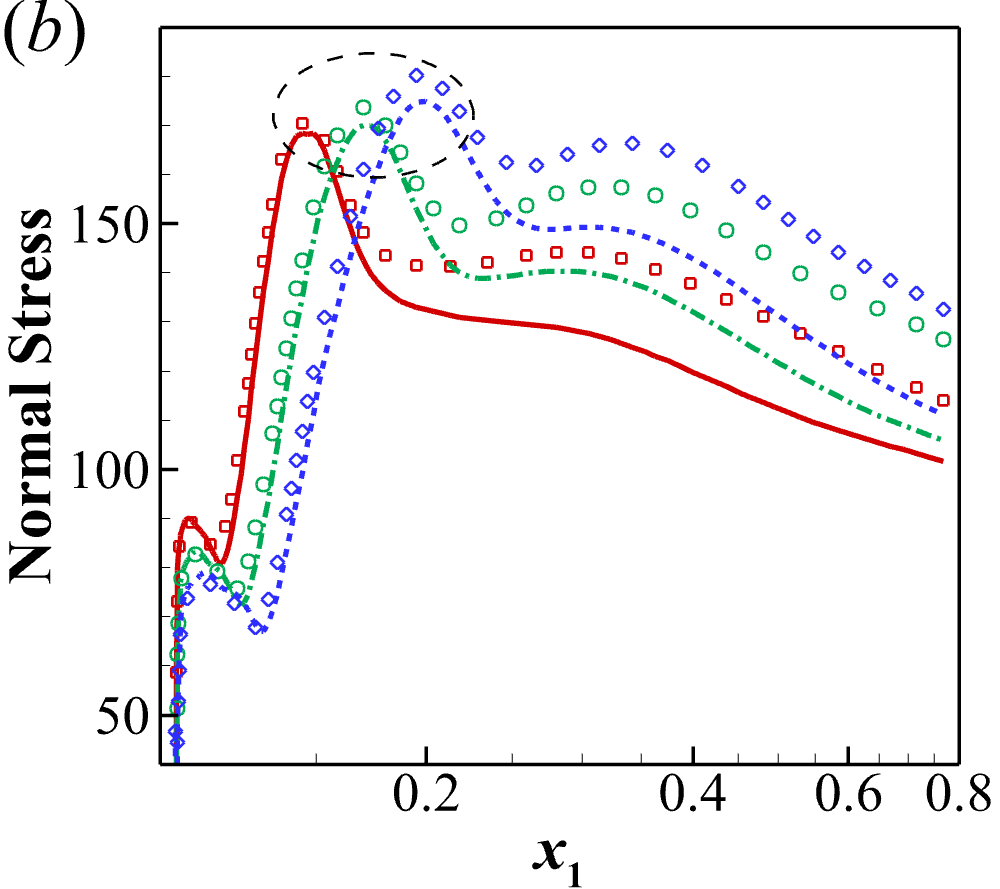}}
    {\includegraphics[width=0.32\linewidth]{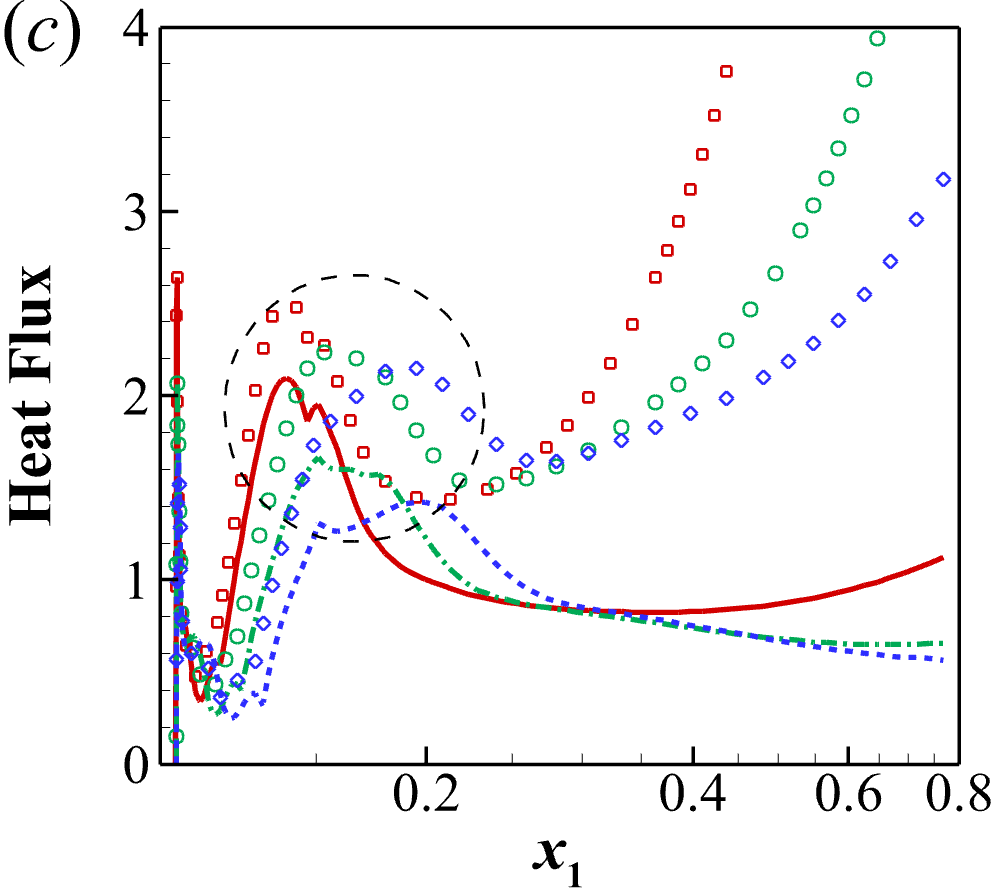}}
    % {\includegraphics[width=0.32\linewidth]{Artworks/JetCFx170.png}}
    % {\includegraphics[width=0.32\linewidth]{Artworks/JetCFx270.png}}
    % {\includegraphics[width=0.32\linewidth]{Artworks/JetCMx370.png}}
    \caption{Stress and heat flux on the lower surface of SLE at different jet PR. 
    The stress are normalized by free flow pressure $p$, while the heat flux is normalized by $p\sqrt{RT}$. 
    % (Second row) Variations of the integrated axial force, normal force and pitch moment coefficients (i.e., $c_a$, $c_n$ and $c_m$) with respect to PR. Force coefficients are forces normalized by $0.5\gamma p\text{Ma}^2\ell$.
    }
    \label{fig:H70tauQFM}
\end{figure}

Figure~\ref{fig:H70tauQFM}(a) shows that the shear stress curves initially oscillate before reattachment (where the downstream counter-clockwise vortex is located), then rise to positive peaks after the shock reattachment, before decreasing further downstream. As the PR increases, positions of the peak of shear stress and the reattachment point shift downstream.
Generally speaking, the GSIS-SST curves follow the same pattern as the GSIS, but consistently lie above the GSIS curves downstream the negative peaks. The discrepancy between GSIS-SST and GSIS increases with PR after reattachment, as highlighted in figure~\ref{fig:H70tauQFM}(a). However, for PR=1.5, the shear stress begins to rise again after $x_1 \approx 0.5~\text{m}$, leading to the largest relative difference of 127\% compared to GSIS.
The distribution of normal stress in figure~\ref{fig:H70tauQFM}(b) follows a similar trend to shear stress, with the maximum relative difference between GSIS-SST and GSIS reaching 20\%.

% and the largest discrepancy between them occurs at the rear of the SLE, where the relative difference reaches 127\%.   figure~\ref{fig:H70tauQFM}(a), while varies little when $x_1>0.3~\text{m}$
% The distribution of normal force is similar to the shear stress, with maximum relative difference between the GSIS-SST and GSIS reaching 20\%. 

% The integrated forces and pitch moment coefficients are compared in the second row of figure~\ref{fig:H70tauQFM}. 
% The axial force coefficients from GSIS-SST are consistently higher than those of GSIS by approximately 0.0013, leading to an increasing relative difference due to the general downward trend as the PR increases. This difference is nearly negligible at $\text{PR}=0$, rises to 13\% at $\text{PR}=1.5$, and reaches approximately 19\% at $\text{PR}=3.5$.
% Meanwhile, GSIS-SST consistently forecasts higher values of normal force and pitch moment than GSIS.
% The relative difference in the normal force remains stable at around 10\% across all PRs, attributed to the concurrent increase in the normal force itself. A similar trend is observed for the pitch moment, with its relative difference hovering within the range of 13\% to 15\%, peaking at $\text{PR}=2.5$.

The surface heat flux curves from GSIS as shown in figure~\ref{fig:H70tauQFM}(c) follow a similar pattern to the shear stress. The GSIS predicts a significantly longer duration of heat flux reduction following the positive peaks, with only the PR=1.5 curve showing a slight recovery in surface heat flux on the downstream half of SLE. In contrast, the GSIS-SST predicts higher positive peaks and a much earlier increase after these peaks, leading to heat flux values that are 4 to 11 times greater than those predicted by GSIS at the rear of SLE. This suggests that the turbulence effects diminish the jet's effectiveness in thermal protection.

\subsection{Turbulent Quantities}

\begin{figure}[h]
    \centering% 0.46 10 -160 60 10 0.49 20 30 140 110
    % \includegraphics[width=0.47\linewidth,trim={10 -150 60 10}, clip]{Artworks/H70MuRCP.png}
    %  \hspace{0.2cm}
    % \includegraphics[width=0.5\linewidth,trim={20 30 140 110}, clip]{Artworks/H70dUdy.png}
    \includegraphics[width=0.95\linewidth,trim={0 0 0 0}, clip]{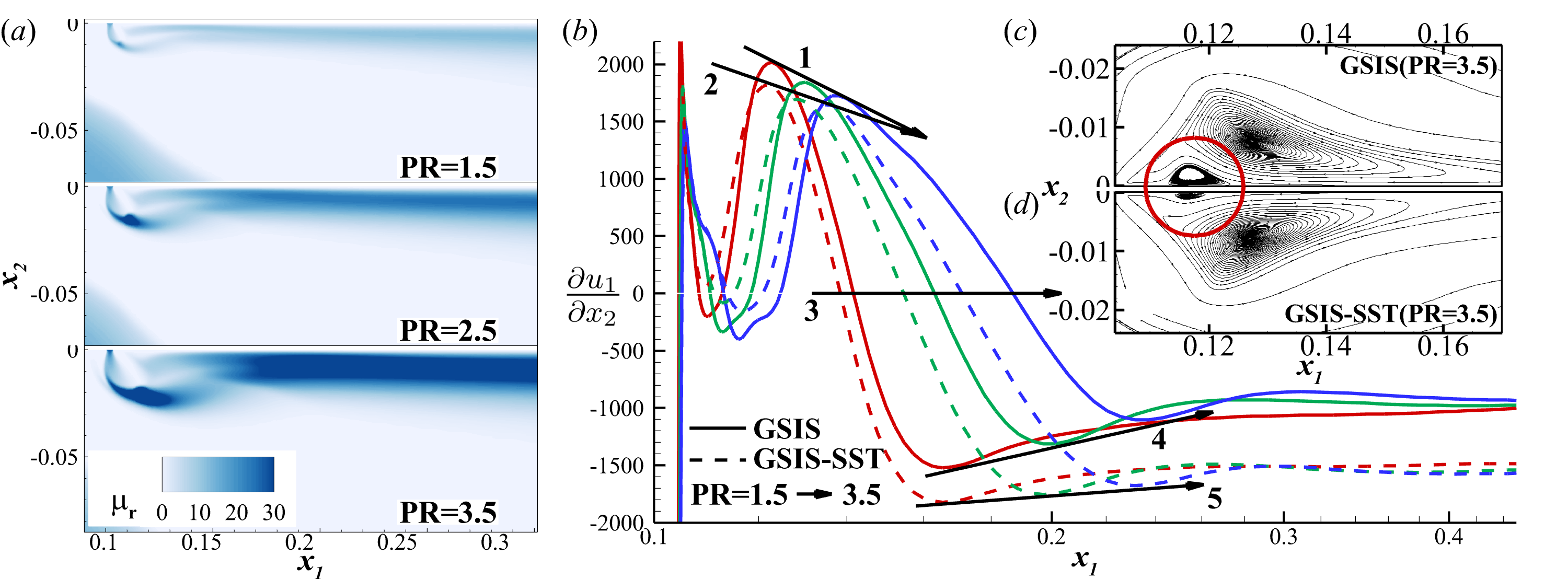}
    \caption{(a) Contours of the turbulent-to-laminar viscosity ratio from GSIS-SST at different PR. 
    (b) Distribution of velocity gradient $\partial u_1/\partial x_2$ on the surface of SLE. The arrows connect key locations of $\sigma_{12}$ across PR. Along the direction of arrow, PR=1.5, 2.5, and 3.5, respectively. (c-d) Streamlines from GSIS (c) and GSIS-SST (d) at PR=3.5. Red circle marks the secondary vortex.
    }
    \label{fig:H70mu_r}
\end{figure}

To understand how the differences between GSIS-SST and GSIS emerge, 
we first examine the turbulent-to-laminar viscosity ratio $\mu_r$, which is pivotal in ascertaining the relative impact of turbulence.
As shown in figure~\ref{fig:H70mu_r}(a), the turbulent jet introduces a high $\mu_r$ into the flow field. This $\mu_r$ decreases in the jet expansion fan but increases across the barrel shock and Mach disk. Notably, high $\mu_r$ persists downstream, after the jet passes the reattachment shock and reattaches to the model surface. The strip-like structures of elevated $\mu_r$ that extend downstream are responsible for the turbulence-related differences between the GSIS and GSIS-SST. As expected, increased value of PR results in a stronger jet, leading to more intense expansion, Mach disk formation, and reattachment shock, all of which generate greater turbulence and higher $\mu_r$.

Before reattachment, the shear stress curves in figure~\ref{fig:H70tauQFM}(a) show a reduction in the magnitude of the negative peaks as PR increases in GSIS, with turbulence further diminishing the peak magnitudes in GSIS-SST. 
The effect of PR is evident in the large downstream vortex shown in figure~\ref{fig:H70TtComment}(b), which expands as PR increases. Concurrently, the velocity gradient $\partial u_1/\partial x_2$ exhibits a lower maximum (arrow 1-2, locations of $\sigma_{12}$ negative peaks) but an extended region where $\partial u_1/\partial x_2>0$ (before arrow 3, locations of $\sigma_{12}$ reattachment points), as seen in figure~\ref{fig:H70mu_r}(b).
The turbulence modeled in GSIS-SST plays a key role in stabilizing the large vortex, preventing vortex breakdown. This stabilization is reflected in the smaller secondary vortex at $x_1 \approx 0.115~\text{m}$ in GSIS-SST compared to GSIS (red circle in figures~\ref{fig:H70mu_r}(c-d)), as the pure GSIS is laminar solver and more prone to vortex breakdown.
Examine the curves from GSIS-SST and GSIS in figure~\ref{fig:H70mu_r}(b), it is seen that the GSIS-SST curves intersect the GSIS curves near arrow 2 within the flow separation region. Upstream of this intersection, the results from GSIS-SST are larger than those from GSIS, while downstream of the intersection, GSIS-SST yields smaller values than GSIS. Additionally, at the negative peaks of $\sigma_{12}$ (arrow 1-2), GSIS-SST predicts a smaller magnitude for $\partial u_1/\partial x_2$ than GSIS. 
The transition from a more skewed curve shape in GSIS to a more Gaussian-like shape in GSIS-SST illustrates the vortex-stabilizing effect of turbulence in the large vortex. Moreover, the reduction in peak values (arrow 2) in GSIS-SST indicates additional turbulent dissipation of kinetic energy due to turbulent viscosity. These changes ultimately explain the differences observed in $\sigma_{12}$ before reattachment.

After reattachment points of $\sigma_{12}$ at $x_1>0.15~\text{m}$ in figure~\ref{fig:H70tauQFM}(a), the difference caused by turbulence becomes more pronounced with increasing PR.
The locations of the positive peaks are indicated in figure~\ref{fig:H70mu_r}(b) by arrow 4 (GSIS) and arrow 5 (GSIS-SST). The directions of arrows 4 and 5 illustrate that the magnitude of $\partial u_1/\partial x_2$ decreases with increasing PR, which explains the reduction in the positive peaks of $\sigma_{12}$.
The difference between arrows 4 and 5 suggests that $\partial u_1/\partial x_2$ in GSIS-SST is consistently larger in magnitude, and this discrepancy scales with PR, leading to an amplified difference in $\sigma_{12}$ between GSIS-SST and GSIS.

As PR increases, the peaks of surface heat flux decrease due to the larger cool jet flow and the enlarged cool separation vortex. The cool vortex further separates the hot free-stream flow from the wall, creating a cooler environment near the surface. Meanwhile, turbulence plays a dual role: it dissipates more kinetic energy into thermal energy, raising the temperature within the vortex, while also transferring more energy from the hot region to the cooler areas through turbulent diffusion, resulting in higher surface heat flux. Since turbulence production intensifies with increasing PR, these effects are also amplified.

\begin{figure}[h]
    \centering
    {\includegraphics[width=0.32\linewidth]{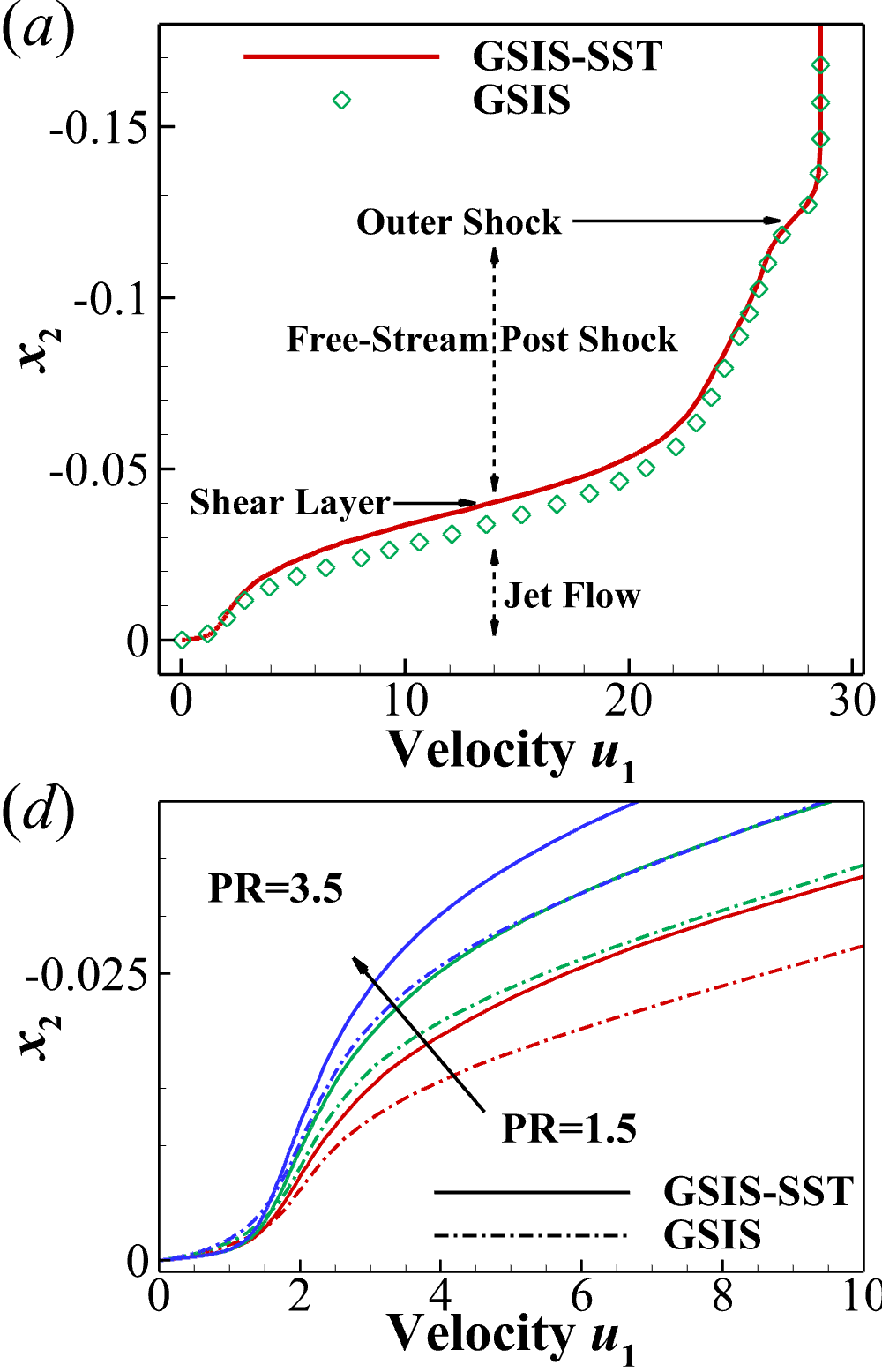}}
    \includegraphics[width=0.32\linewidth]{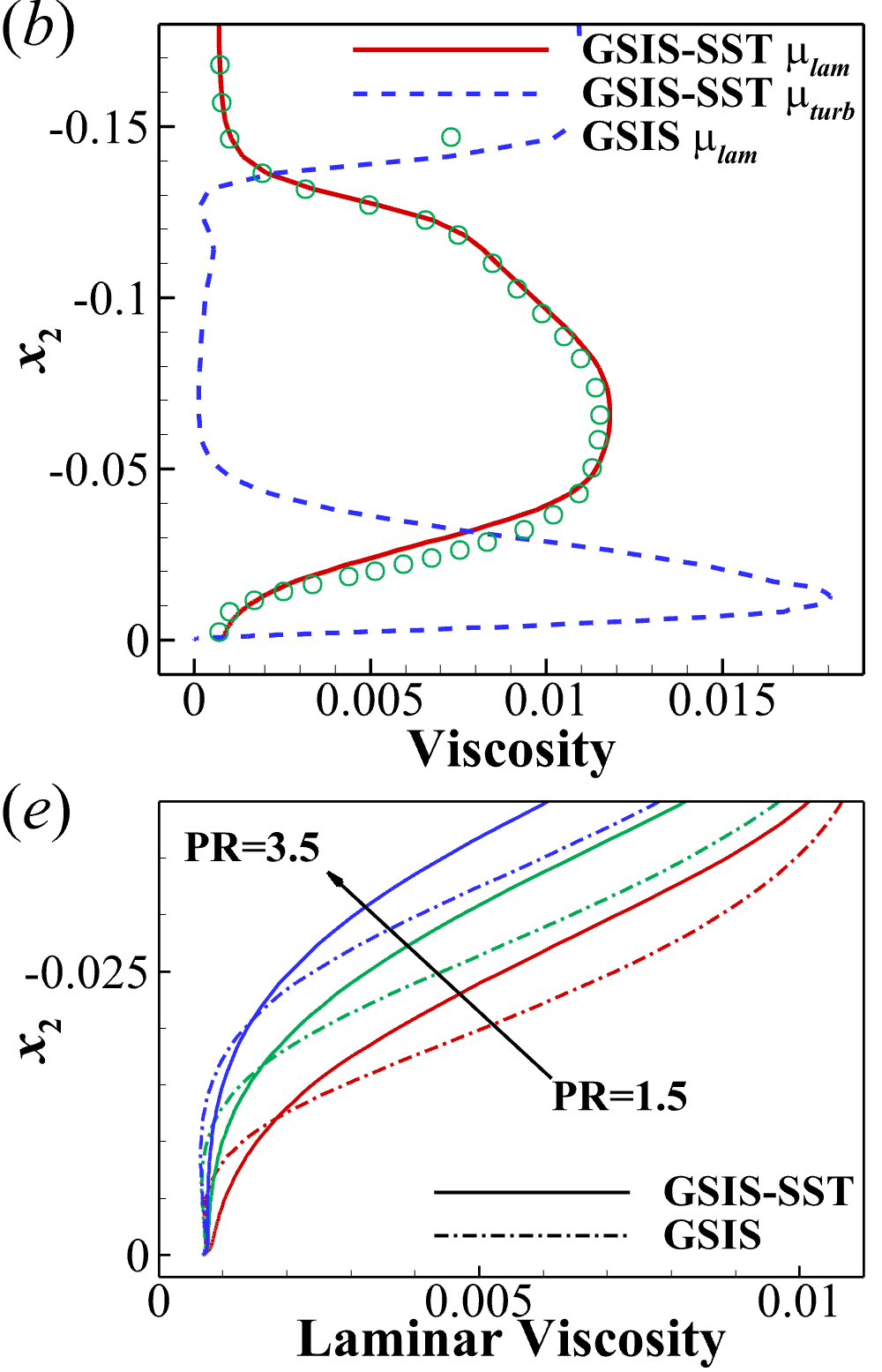}
    \includegraphics[width=0.32\linewidth]{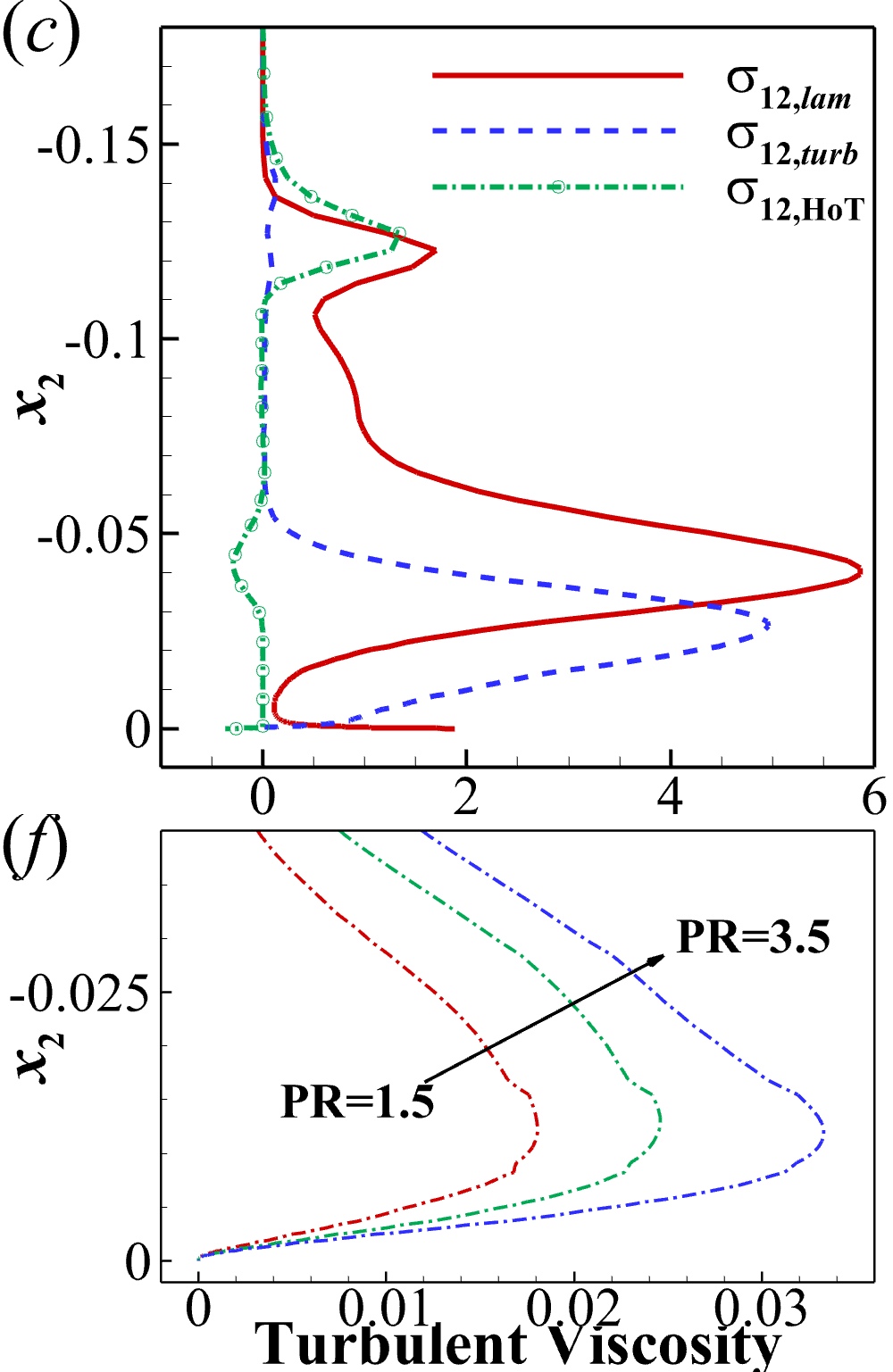}
    \caption{(First row) Macroscopic flow quantities sampled at $x_1=0.35~\text{m}$, when PR=1.5. The velocity, viscosity and shear stress are normalized by $\sqrt{RT}$, $pL/\sqrt{RT}$ and $p$, respectively. %\leir{remove the Mach number; in the order of velocity, viscosity, stress }
    (Second row) Detailed profiles of velocity, laminar and turbulent viscosities in the near wall region at $x_1=0.35~\text{m}$ with different PR ratios. Turbulent viscosity is extracted from GSIS-SST only.
    }
    \label{fig:H70Vis}
\end{figure}

To further investigate how turbulence and PR influence the flow field after reattachment, we plot the velocity, shear stress, and viscosity at $x_1 = 0.35~\text{m}$ for PR = 1.5 in figure~\ref{fig:H70Vis}, where the jet has reattached to the model surface and a typical boundary layer has formed. Distinct flow layers are marked in figure~\ref{fig:H70Vis}(a), and the differences between GSIS-SST and GSIS begin to emerge around the indistinct shear layer at $x_2 \approx -0.05~\text{m}$. This is because $\mu_r$ is only significant near the wall, as shown by the viscosity profiles in figure~\ref{fig:H70Vis}(b).

The constitutive relation in GSIS-SST includes contributions from the laminar part $\sigma_{12,lam}$, turbulent $\sigma_{12,turb}$, and rarefied part $\sigma_{12,\text{HoT}}$~\citep{Tian2024Multiscale}. As shown in figure~\ref{fig:H70Vis}(c), the laminar part, $\sigma_{12,lam}$, peaks at the outer shock and the shear layer, driven by the high velocity gradients across these regions, as well as the increase in post-shock temperature (and hence the laminar viscosity $\mu_{lam}$, as shown in figure~\ref{fig:H70Vis}(b)). A similar high velocity gradient near the wall also produces $\sigma_{12,lam}$, despite the reduction in $\mu_{lam}$ as the flow approaches the colder wall.
At the first peak ($x_2 \approx -0.125~\text{m}$), the gas rarefaction effect dominates over turbulence. Near the shear layer ($x_2 \approx -0.05~\text{m}$), the turbulent contribution becomes comparable to the laminar part, while the rarefaction term, $\sigma_{12,\text{HoT}}$, exerts a small negative effect on the shear stress. 
At the last peak of $\sigma_{12,lam}$ near wall, turbulent part drops to zero due to the disappearance of $\mu_{turb}$ enforced by the wall boundary condition, while the rarefaction effects again exhibit negative value in the Knudsen layer: in rarefied gas flows, the rarefaction near the Knudsen layer results in an effective viscosity that is smaller than the molecular viscosity. Consequently, $\sigma_{12,\text{HoT}}$ and $\sigma_{12,lam}$ have opposite signs~\citep{Su20217PRE}. 
% It is seen here that the outer part of shear layer experiences mutual effect of rarefaction and turbulence, which acts oppositely.

As PR increases, the jet flow layer becomes thicker, and the profiles of $u_1$ and $\mu_{lam}$ shift significantly in the $x_2$ direction, taking a larger distance to reach the mainstream levels, as shown in figures~\ref{fig:H70Vis}(d) and (e). The wall-normal gradients at the wall decrease, leading to a reduction in both $\sigma_{12}$ and the heat flux $q$ from GSIS at higher PR, as depicted in figure~\ref{fig:H70tauQFM}.

% (figure~\ref{fig:H70Vis}(d))
In time-averaged flow fields with a cold wall, it is well-known that the turbulent velocity and temperature profiles modeled in turbulence simulations typically exhibit greater thickness and steeper normal gradients at the wall~\citep{Schlichting2017Boundary}. In contrast, the normal gradients at wall are smaller in the thinner laminar profiles. This behavior is evident in figures~\ref{fig:H70Vis}(d) and (e), where $\mu_{lam} \propto T^{0.74}$ for nitrogen.
Comparing figure~\ref{fig:H70Vis}(f) with (d) and (e), it is evident that the largest difference in the slopes of $u_1$ and $\mu_{lam}$ between GSIS-SST and GSIS occurs near the peaks of $\mu_{turb}$. The presence of $\mu_{turb}$ represents turbulent diffusion, which enhances momentum transfer from high-velocity to low-velocity regions, and similarly facilitates energy transfer.
As a result, the shear stress and heat flux distributions in figure~\ref{fig:H70tauQFM} from GSIS-SST are consistently higher than those from GSIS after reattachment.

% With enhanced momentum exchange by turbulence, the velocity distribution exhibits reduced slopes as per comparison between GSIS-SST and GSIS in figure~\ref{fig:H70Vis}(d), especially at high $\mu_{turb}$ region (figure~\ref{fig:H70Vis}(f)). The enhanced energy exchange by turbulence could be observed through $\mu_{lam}$ in figure~\ref{fig:H70Vis}(e), with the GSIS-SST solution predicts an earlier and smoother drop of $\mu_{lam}$. 
% Since turbulence smooths the momentum and energy change only inside the flow field, gradients adjacent to the wall go up (similar as the analysis of positive peaks of $\sigma_{12}$), and wall slip effect is slightly intensified. Meanwhile, increase in PR thickens jet flow layer, yet reducing the gradients of velocity near wall. Due to the opposite effect of turbulence and PR, the turbulence shifts of shear stress and normal force thus suffer little from increased PR in the middle part of the model. 

 % It is observed that turbulence also slightly promotes wall slip, leading a higher $\mu_{lam}$ at wall.
% \begin{figure}
%     \centering
%     {\includegraphics[width=0.32\linewidth]{Artworks/H70PRVelDetail.png}}
%     {\includegraphics[width=0.32\linewidth]{Artworks/H70PRVis.png}}
%     {\includegraphics[width=0.32\linewidth]{Artworks/H70PRVisTurb.png}}
%     \caption{Detailed profiles of viscosity and velocity in the near wall region at $x_1=0.35~\text{m}$ with different PR ratios. Turbulent viscosity is extracted from GSIS-SST only. }
%     \label{fig:H70VisDetail}
% \end{figure}

% \leib{rise of the curve near the end}
As the jet flow progresses toward the end of the SLE, turbulent diffusion continues to enhance momentum and energy transfer in the wall-normal direction. However, the $\mu_{turb}$ profiles in figure~\ref{fig:H70Vis}(f) show that while the magnitude of $\mu_{turb}$ increases with PR, the location of its peak does not shift as significantly as the profiles of $u_1$ and $\mu_{lam}$. In other words, the thickness of the dominant $\mu_r$ layer remains relatively constant across different PRs. This suggests that at lower PR, the higher momentum and energy in the post-shock free flow are closer to the turbulence-dominated layer.

Consequently, the cool near-wall jet flow is more easily heated and accelerated at lower PR, leading to an earlier onset of stress/heat flux reduction, as evidenced by the rise in GSIS-SST curves near the end of the SLE.
\section{The influence of AoA}\label{sec:AoA}

The AoA is adjusted to 0°, 15° and 30° to simulate the maneuvers of hypersonic vehicles. The flight altitude is increased to 80~km, since attitude adjustment of hypersonic flight vehicles is more efficient at higher altitudes. Referring to \cite{United1976United}, the free flow pressure is $p=1.05$ Pa, temperature is $T=198.6$ K, viscosity is $\mu=1.32\times 10^{-5}~\text{Pa}\cdot \text{s}$, so the Knudsen number is 0.0375 and Reynolds numbers is 987. The jet PR is kept at 2.5.

\subsection{Turbulence influence on surface quantities}

\begin{figure}
    \centering
    \includegraphics[width=0.4\linewidth,trim={0 0 0 0},clip]{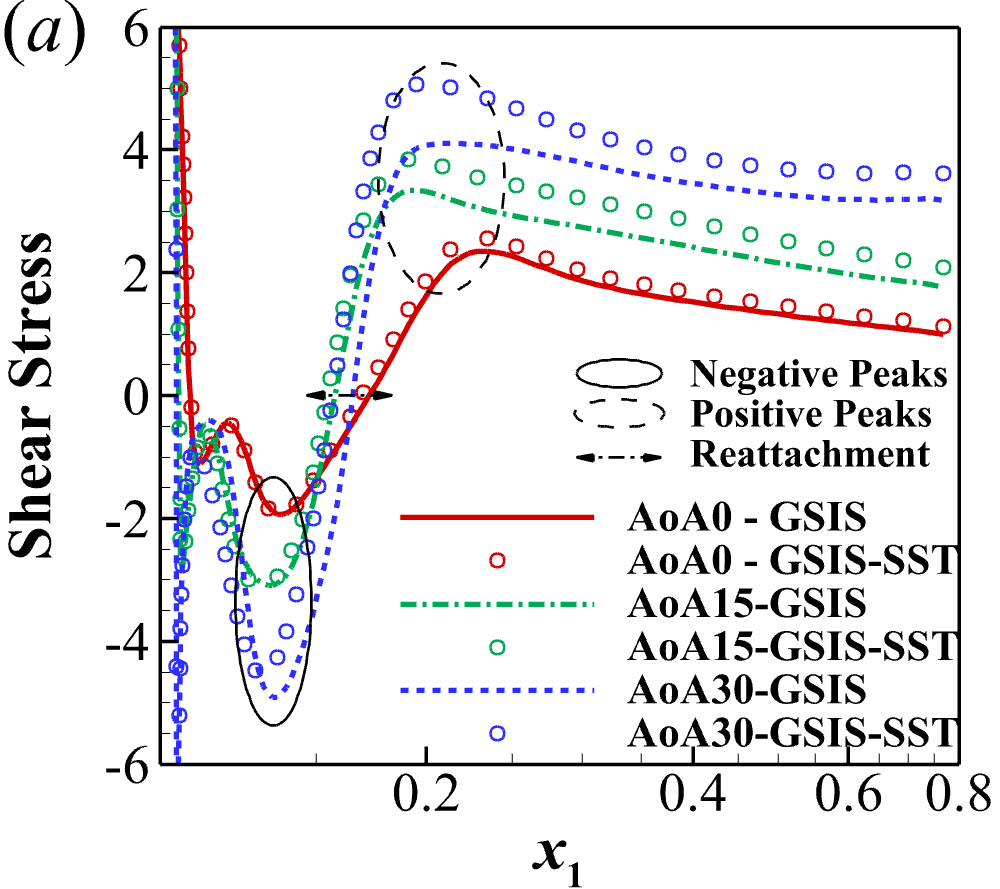}\hspace{0.5cm}
    \includegraphics[width=0.4\linewidth,trim={0 0 0 0},clip]{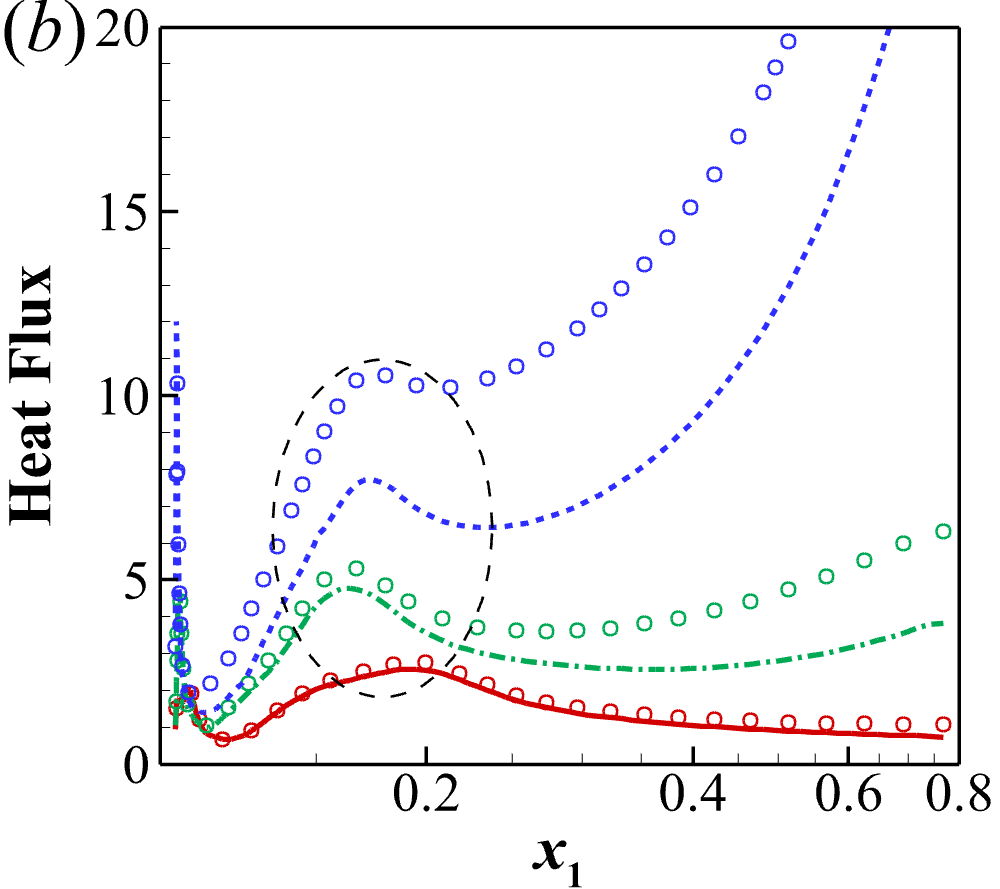}
    \caption{Variation of surface shear stress and heat flux with AoA.
    }
    \label{fig:H80tauxy}
\end{figure}

% \leib{First describe the phenomena, then analysis the difference!}
The shear stress depicted in figure~\ref{fig:H80tauxy}(a) exhibits a similar trend to the results at 70~km (figure~\ref{fig:H70tauQFM}). A negative peak in shear stress is observed around $x_1 = 0.15~\text{m}$. Subsequently, the curves rise to a positive peak near $x_1 \approx 0.2~\text{m}$ and then decline continuously. The magnitude of both the negative and positive peaks in shear stress increases with the AoA. 
The surface heat flux shown in figure~\ref{fig:H80tauxy}(b) exhibits a local peak near $x_1 = 0.16\sim0.20$~m across all AoAs. Following this peak, the heat flux decreases continuously at AoA = 0°, but it increases at downstream when AoA = 15° and 30°. The GSIS-SST solutions predict higher heat flux than GSIS, indicating that turbulence enhances mixing and reduces the cooling effect. Such discrepancy also scales apparently with AoA.
% The maximum relative difference between GSIS-SST and GSIS can reach up to 70\%, which its location shifts upstream with increasing AoA.

%The turbulence induced discrepancy increases with AoA. %At such high altitudes, the effect of turbulence on surface stress is minimal, resulting in the integrated forces and moments deviating by less than 5\% between the GISS-SST and GSIS models.
% Positive peak 23 A30; 

% Despite the less than 5\% relative difference in pitch moment as shown in figure~\ref{fig:Mx3}, the effect of turbulence on surface shear stress and heat flux remains significant. 

\subsection{Turbulent Quantities}

The distribution of the turbulent-to-laminar viscosity ratio $\mu_r$ follows similar pattern as at 70~km (figure~\ref{fig:H70mu_r}(a)), with high $\mu_{r}$ still concentrates around Mach disk, downstream vortex and near-wall jet layer. As AoA increases, the downstream vortex is compressed in the $x_2$ direction while being elongated in the $x_1$ direction, and the wall-normal velocity gradient steepens, promoting turbulence generation.
The produced turbulence leads to similar discrepancies around negative and positive peaks of shear stress (figure~\ref{fig:H80tauxy}(a)) as at 70~km, with a positive relation between turbulence-induced difference and AoA around the peaks.

\begin{figure}%[h]
    \centering
    % \vspace{0.3cm}
    % \subfloat[AoA= 0°]
    {\includegraphics[width=0.24\linewidth]{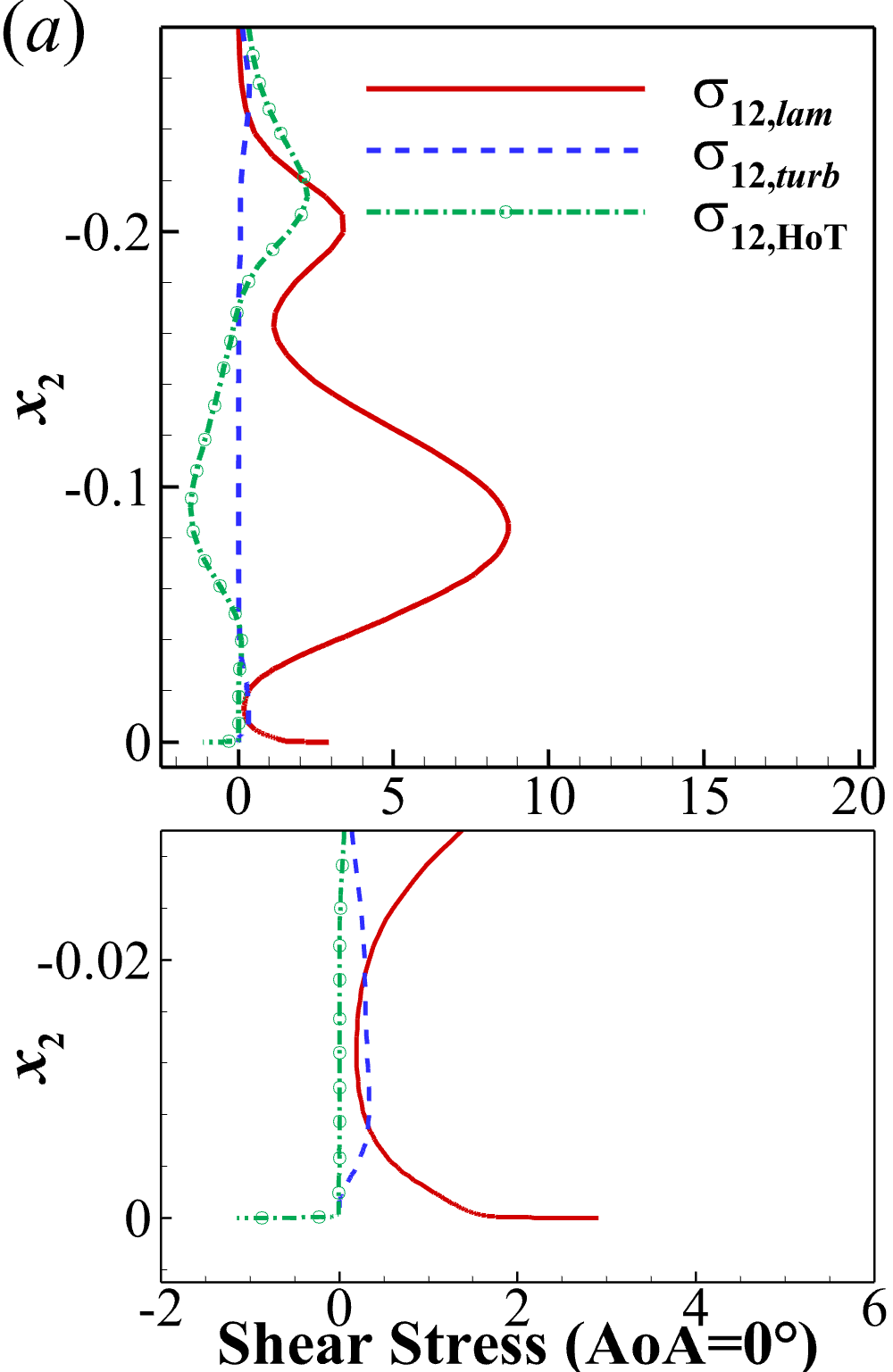}}
    % \subfloat[AoA= 15°]
    {\includegraphics[width=0.24\linewidth]{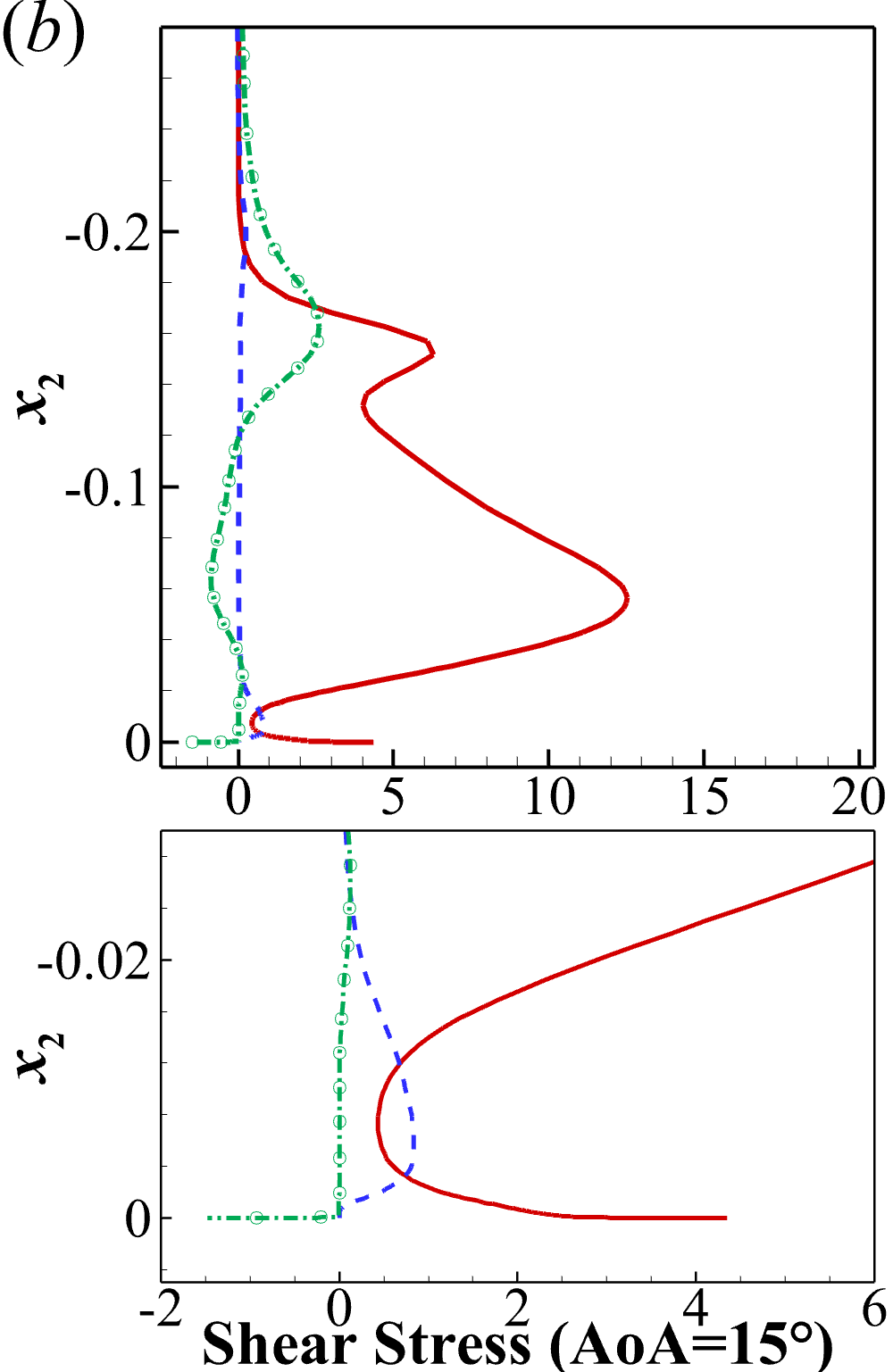}}
    % \subfloat[AoA= 30°]
    {\includegraphics[width=0.24\linewidth]{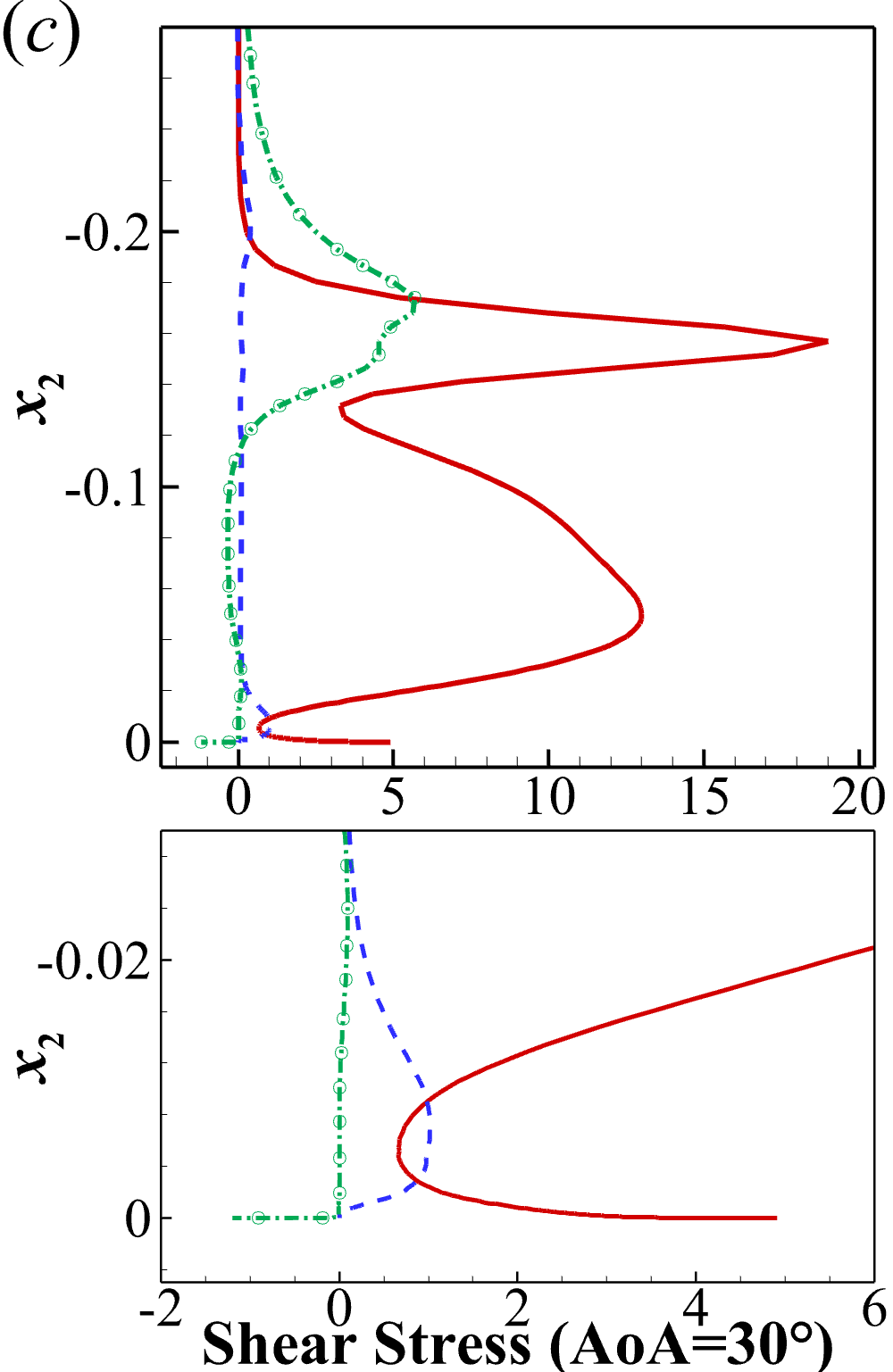}}
    % \subfloat[Viscosities]
    {\includegraphics[width=0.24\linewidth]{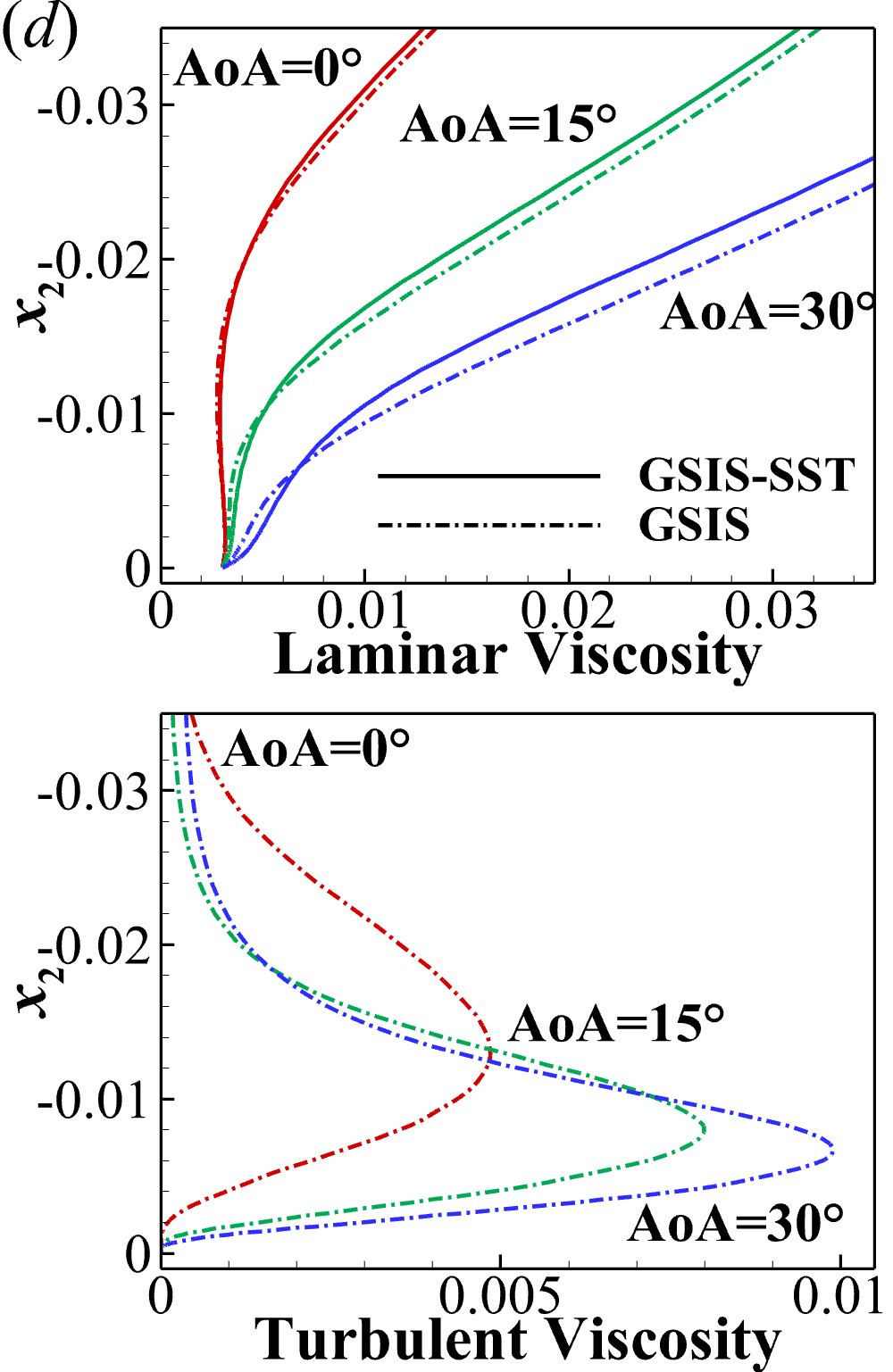}}
    \caption{(a-c) GSIS-SST results of the NS laminar, turbulent and rarefied parts of shear stress at $x_1=0.35~\text{m}$. (d) Variation of viscosities with AoA.
    %(Second row) Profiles of laminar, turbulent viscosities and their ratio at $x_1=0.35~\text{m}$ at altitude 80~km. Turbulent viscosity and viscosity ratio are extracted from GSIS-SST only.
    }
    \label{fig:H80Proftxy}
\end{figure}

Figure~\ref{fig:H80Proftxy}(a-c) compares the profiles of shear stress and viscosity between GSIS-SST and GSIS, at $x_1 = 0.35~\text{m}$. The shear stress profiles show a compression toward the wall as AoA increases, with regions dominated separately by turbulent contributions and rarefaction-induced HoT part.
Turbulent part $\sigma_{12,turb}$ is negligible except within the jet boundary layer, where it matches the magnitude of $\sigma_{12,lam}$ and increases with AoA, see the enlarged view in the second row of figure~\ref{fig:H80Proftxy}. 
Rarefaction-induced shear stress $\sigma_{12,HoT}$ extends further than at 70~km, but diminishes in the post-shock free-flow region as the AoA increases due to higher density and reduced rarefaction effects. Among the three AoAs, the magnitude of $\sigma_{12,HoT}$ at wall is largest when AoA = 15°, suggesting a complex AoA-dependent relationship. 

In figure~\ref{fig:H80Proftxy}(d), both $\mu_{lam}$ and $\mu_{turb}$ increase near the wall with increasing AoA. The rise in $\mu_{lam}$ is due to thinner jet flow layers and higher temperature gradients. GSIS-SST’s $\mu_{turb}$ intensifies these gradients, with peak values rising with AoA due to enhanced compression. However, the range of turbulent influence narrows significantly, especially when AoA increases from 0° to 15°, indicating AoA’s role in modulating the balance between the magnitude and range of $\mu_r$ at fixed PR. 
Nonetheless, discrepancies between GSIS-SST and GSIS in shear stress and heat flux increase with AoA.

\section{Conclusions}\label{sec:conclusion}

We have conducted an analysis of lateral turbulent jets in rarefied flows using the multiscale GSIS-SST method. This method converges to the conventional RANS solver for turbulence at lower altitudes and to the Boltzmann solver for rarefied gas flows at higher altitudes. 
The GSIS-SST solver is valid across all altitudes and effectively manages scenarios where both turbulence and gas rarefaction effects are present. 
Specifically, within the altitude range of 50 to 80 km, both turbulent and rarefaction effects are significant and cannot be adequately captured by either the RANS solver or the Boltzmann equation alone. In the rarefied environment, the presence of turbulent jet affects the near wall velocity and temperature gradients, through enhanced mixing in the flow field and the formation of turbulent-like boundary layer. When both turbulence and gas rarefaction are considered, 
turbulence induced difference in stress and heat flux expands with PR before the reduction effect degrades, while the degradation of reduction effect appears earlier at lower PR, leading to over 1 (11) times greater shear stress (surface heat flux) at then end of the SLE, at an altitude of 70 km, AoA = 15° and PR=1.5.
Variation in angle of attack modulates the balance between intensity and range of turbulent viscosity, resulting in enlarged difference in shear stress and heat flux induced by turbulence as AoA increases.

This research offers new insights into the intricate flow dynamics of lateral jets in high-altitude applications, emphasizing the critical interplay between turbulent and rarefied effects.
Leveraging these insights, future efforts in the design and optimization of jet-based active flow control systems can lead to more reliable solutions for hypersonic vehicles.

\vspace{0.3cm}
\noindent \textbf{Acknowledgments}. This work is supported by the National Natural Science Foundation of China (12450002) and the Stable Support Plan (80000900019910072348). Special thanks are given to the Center for Computational Science and Engineering at the Southern University of Science and Technology. \\  % 

\noindent \textbf{Declaration of interests}. The authors report no conflict of interest.

%\noindent This work is supported by the National Natural Science Foundation of China (12450002). The authors report no conflict of interest.

\bibliographystyle{jfm}
\bibliography{LJ}

\begin{thebibliography}{28}
\expandafter\ifx\csname natexlab\endcsname\relax\def\natexlab#1{#1}\fi
\def\au#1{#1} \def\ed#1{#1} \def\yr#1{#1}\def\at#1{#1}\def\jt#1{\textit{#1}}
  \def\bt#1{#1}\def\bvol#1{\textbf{#1}} \def\vol#1{#1} \def\pg#1{#1}
  \def\publ#1{#1}\def\arxiv#1{#1}\def\org#1{#1}\def\st#1{\textit{#1}}

\bibitem[Aristov(2001)]{Aristov2001Direct}
{\sc \au{Aristov, V.V.}} \yr{2001} {\em Direct Methods for Solving the
  Boltzmann Equation and Study of Nonequilibrium Flows\/}.  \publ{Springer}.

\bibitem[Bird(1994)]{Bird1994Molecular}
{\sc \au{Bird, G.~A.}} \yr{1994} {\em Molecular Gas Dynamics and the Direct
  Simulation of Gas Flows\/}.  \publ{Oxford University Press Inc, New York:
  Oxford Science Publications}.

\bibitem[Boles {\em et~al.\/}(2010)Boles, Edwards \& Baurle]{Boles2010Large}
{\sc \au{Boles, J.~A.}, \au{Edwards, J.~R.} \& \au{Baurle, R.~A.}} \yr{2010}
  \at{Large-eddy/{Reynolds-Averaged Navier-Stokes} simulations of sonic
  injection into {Mach} 2 crossflow}.  \jt{AIAA J.}  \bvol{48},
  \pg{1444--1456}.

\bibitem[Boyd {\em et~al.\/}(1995)Boyd, Chen \& Candler]{Boyd1995Predicting}
{\sc \au{Boyd, I.~D.}, \au{Chen, G.} \& \au{Candler, G.~V.}} \yr{1995}
  \at{Predicting failure of the continuum fluid equations in transitional
  hypersonic flows}.  \jt{Phys. Fluids}  \bvol{7},  \pg{210--219}.

\bibitem[Chapman \& Cowling(1990)]{Chapman1990Mathematical}
{\sc \au{Chapman, S.} \& \au{Cowling, T.~G.}} \yr{1990} {\em The Mathematical
  Theory of Non-uniform Gases\/}.  \publ{Cambridge University Press}.

\bibitem[Fei(2023)]{Fei2023JCP}
{\sc \au{Fei, F.}} \yr{2023}  \at{{A time-relaxed Monte Carlo method preserving
  the Navier-Stokes asymptotics}}.  \jt{J. Comput. Phys.}  \bvol{486},
  \pg{112128}.

\bibitem[Gallis {\em et~al.\/}(2017)Gallis, Bitter, Koehler, Torczynski,
  Plimpton \& Papadakis]{Gallis2017Molecular}
{\sc \au{Gallis, M.~A.}, \au{Bitter, N.~P.}, \au{Koehler, T.~P.},
  \au{Torczynski, J.~R.}, \au{Plimpton, S.~J.} \& \au{Papadakis, G.}} \yr{2017}
   \at{Molecular-level simulations of turbulence and its decay}.  \jt{Phys.
  Rev. Lett.}  \bvol{118},  \pg{064501}.

\bibitem[Gimelshein {\em et~al.\/}(2002)Gimelshein, Alexeenko \&
  Levin]{Gimelshein2002Modeling}
{\sc \au{Gimelshein, S.~F.}, \au{Alexeenko, A.~A.} \& \au{Levin, D.~A.}}
  \yr{2002}  \at{Modeling of the interaction of a side jet with a rarefied
  atmosphere}.  \jt{Journal of Spacecraft and Rockets}  \bvol{39},
  \pg{168--176}.

\bibitem[Goldberg \& Apsley(1997)]{Goldberg1997wall}
{\sc \au{Goldberg, U.} \& \au{Apsley, D.}} \yr{1997}  \at{A wall-distance-free
  low $\text{Re}$ $k-\epsilon$ turbulence model}.  \jt{Computer Methods in
  Applied Mechanics and Engineering}  \bvol{145},  \pg{227--238}.

\bibitem[Gorji {\em et~al.\/}(2011)Gorji, Torrilhon \& Jenny]{Gorji2011JFM}
{\sc \au{Gorji, M.~H.}, \au{Torrilhon, M.} \& \au{Jenny, P.}} \yr{2011}
  \at{{Fokker–Planck model for computational studies of monatomic rarefied
  gas flows}}.  \jt{J. Fluid Mech.}  \bvol{680},  \pg{574--601}.

\bibitem[Karpuzcu \& Levin(2023)]{Karpuzcu2023Study}
{\sc \au{Karpuzcu, I.~T.} \& \au{Levin, D.~A.}} \yr{2023}  \at{Study of
  side-jet interactions over a hypersonic cone flow using kinetic methods}.
  \jt{AIAA J.}  \bvol{61},  \pg{4741--4751}.

\bibitem[Li {\em et~al.\/}(2021)Li, Zeng, Su \& Wu]{li2021uncertainty}
{\sc \au{Li, Q.}, \au{Zeng, J.~N.}, \au{Su, W.} \& \au{Wu, L.}} \yr{2021}
  \at{Uncertainty quantification in rarefied dynamics of molecular gas: rate
  effect of thermal relaxation}.  \jt{J. Fluid Mech.}  \bvol{917},  \pg{A58}.

\bibitem[Liu {\em et~al.\/}(2020)Liu, Zhu \& Xu]{Liu2020Unified}
{\sc \au{Liu, C.}, \au{Zhu, Y.~J.} \& \au{Xu, K.}} \yr{2020}  \at{Unified
  gas-kinetic wave-particle methods {I}: Continuum and rarefied gas flow}.
  \jt{J. Comput. Phys.}  \bvol{401},  \pg{108977}.

\bibitem[Liu {\em et~al.\/}(2024)Liu, Zhang, Zeng \& Wu]{Liu2024Further}
{\sc \au{Liu, W.}, \au{Zhang, Y.~B.}, \au{Zeng, J.~N.} \& \au{Wu, L.}}
  \yr{2024}  \at{Further acceleration of multiscale simulation of rarefied gas
  flow via a generalized boundary treatment}.  \jt{J. Comput. Phys.}
  \bvol{503},  \pg{112830}.

\bibitem[Mahesh(2013)]{Mahesh2013Interaction}
{\sc \au{Mahesh, K.}} \yr{2013}  \at{The interaction of jets with crossflow}.
  \jt{Annu. Rev. Fluid Mech.}  \bvol{45},  \pg{379--407}.

\bibitem[Menter(1994)]{Menter1994Two}
{\sc \au{Menter, F.~R.}} \yr{1994}  \at{Two-equation eddy-viscosity turbulence
  models for engineering applications}.  \jt{AIAA J.}  \bvol{32},
  \pg{1598--1605}.

\bibitem[Miller {\em et~al.\/}(2018)Miller, Medwell, Doolan \&
  Kim]{Miller2018Transient}
{\sc \au{Miller, W.~A.}, \au{Medwell, P.~R.}, \au{Doolan, C.~J.} \& \au{Kim,
  M.}} \yr{2018}  \at{Transient interaction between a reaction control jet and
  a hypersonic crossflow}.  \jt{Phys. Fluids}  \bvol{30},  \pg{046102}.

\bibitem[NOAA {\em et~al.\/}(1976)NOAA, NASA \& {U.S.A.F}]{United1976United}
{\sc \au{NOAA}, \au{NASA} \& \au{{U.S.A.F}}} \yr{1976} {\em U.S. Standard
  Atmosphere, 1976\/}. {\em NOAA - SIT\/} 76-1562.  \publ{U.S. Government
  Printing Office}.

\bibitem[Rowton {\em et~al.\/}(2024)Rowton, Medwell \&
  Chin]{Rowton2024ANumerical}
{\sc \au{Rowton, Harry~C.}, \au{Medwell, Paul~R.} \& \au{Chin, Rey}} \yr{2024}
  \at{A numerical study of the effects of jet-aft wall temperatures on the
  dynamics of jets in hypersonic crossflows}.  \jt{Phys. Fluids}  \bvol{36},
  \pg{016140}.

\bibitem[Sanaka {\em et~al.\/}(2024)Sanaka, Sharma, Ramana~Murty \&
  Durga~Rao]{Sanaka2024Re}
{\sc \au{Sanaka, S.~P.}, \au{Sharma, R.~K.}, \au{Ramana~Murty, G.~V.} \&
  \au{Durga~Rao, K.}} \yr{2024}  \at{Re-entry vehicle performance analysis
  under the control of lateral jet}.  \jt{The Aeronautical Journal}
  \bvol{128},  \pg{756--770}.

\bibitem[Schlichting \& Gersten(2017)]{Schlichting2017Boundary}
{\sc \au{Schlichting, Hermann} \& \au{Gersten, Klaus}} \yr{2017} {\em
  Boundary-Layer Theory\/}.  \publ{Berlin, Heidelberg: Springer Berlin
  Heidelberg}.

\bibitem[Su {\em et~al.\/}(2017)Su, Lindsay, Liu \& Wu]{Su20217PRE}
{\sc \au{Su, W.}, \au{Lindsay, S.}, \au{Liu, H.~H.} \& \au{Wu, L.}} \yr{2017}
  \at{{Comparative study of the discrete velocity and lattice Boltzmann methods
  for rarefied gas flows through irregular channels}}.  \jt{Phys. Rev. E}
  \bvol{96},  \pg{023309}.

\bibitem[Su {\em et~al.\/}(2020{\natexlab{{\em a\/}}})Su, Zhu, Wang, Zhang \&
  Wu]{Su2020Can}
{\sc \au{Su, W.}, \au{Zhu, L.~H.}, \au{Wang, P.}, \au{Zhang, Y.~H.} \& \au{Wu,
  L.}} \yr{2020{\natexlab{{\em a\/}}}}  \at{Can we find steady-state solutions
  to multiscale rarefied gas flows within dozens of iterations?}  \jt{J.
  Comput. Phys.}  \bvol{407},  \pg{109245}.

\bibitem[Su {\em et~al.\/}(2020{\natexlab{{\em b\/}}})Su, Zhu \&
  Wu]{Su2020Fast}
{\sc \au{Su, W.}, \au{Zhu, L.~H.} \& \au{Wu, L.}} \yr{2020{\natexlab{{\em
  b\/}}}}  \at{Fast convergence and asymptotic preserving of the general
  synthetic iterative scheme}.  \jt{SIAM J. Sci. Comput.}  \bvol{42},
  \pg{B1517--B1540}.

\bibitem[Tian \& Wu(2025)]{Tian2024Multiscale}
{\sc \au{Tian, S.~Y.} \& \au{Wu, L.}} \yr{2025}  \at{Multiscale modelling and
  simulation of coexisting turbulent and rarefied gas flows}.  \jt{J. Fluid
  Mech.}  \bvol{1002},  \pg{A10}.

\bibitem[Wu {\em et~al.\/}(2015)Wu, White, Scanlon, Reese \&
  Zhang]{Wu2015kinetic}
{\sc \au{Wu, L.}, \au{White, C.}, \au{Scanlon, T.~J.}, \au{Reese, J.~M.} \&
  \au{Zhang, Y.~H.}} \yr{2015}  \at{A kinetic model of the {Boltzmann equation}
  for non-vibrating polyatomic gases}.  \jt{J. Fluid Mech.}  \bvol{763},
  \pg{24--50}.

\bibitem[Xu \& Huang(2010)]{Xu2010unified}
{\sc \au{Xu, K.} \& \au{Huang, J.~C.}} \yr{2010}  \at{A unified gas-kinetic
  scheme for continuum and rarefied flows}.  \jt{J. Comput. Phys.}  \bvol{229},
   \pg{7747--7764}.

\bibitem[Zhang {\em et~al.\/}(2024)Zhang, Zeng, Yuan, Liu, Li \&
  Wu]{Zhang2024Efficient}
{\sc \au{Zhang, Y.~B.}, \au{Zeng, J.~N.}, \au{Yuan, R.~F.}, \au{Liu, W.},
  \au{Li, Q.} \& \au{Wu, L.}} \yr{2024}  \at{Efficient parallel solver for
  rarefied gas flow using {GSIS}}.  \jt{Comput. Fluids}  \bvol{281},
  \pg{106374}.

\end{thebibliography}

\end{document}